\documentclass[twocolumn]{aastex631}

\usepackage{newtxtext,newtxmath}

\begin{document}

\title{The Physical Origin and Time Lag of Multi-Frequency Flares from Sgr~A$^*$}

\author[0000-0003-0292-2773]{Hong-Xuan Jiang}
\affiliation{Tsung-Dao Lee Institute, Shanghai Jiao Tong University, Shengrong Road 520, Shanghai, 201210, China}
\author[0000-0002-8131-6730]{Yosuke Mizuno}
\affiliation{Tsung-Dao Lee Institute, Shanghai Jiao Tong University, Shengrong Road 520, Shanghai, 201210, China}
\affiliation{School of Physics and Astronomy, Shanghai Jiao Tong University, 
800 Dongchuan Road, Shanghai, 200240, China}
\affiliation{Key Laboratory for Particle Physics, Astrophysics and Cosmology (MOE), Shanghai Key Laboratory for Particle Physics and Cosmology, Shanghai Jiao-Tong University,800 Dongchuan Road, Shanghai, 200240, People's Republic of China}
\affiliation{Institut f\"ur Theoretische Physik, Goethe-Universit\"at Frankfurt, Max-von-Laue-Stra{\ss}e 1, D-60438 Frankfurt am Main, Germany}
\author[0000-0002-4064-0446]{Indu K. Dihingia}
\affiliation{Tsung-Dao Lee Institute, Shanghai Jiao Tong University, Shengrong Road 520, Shanghai, 201210, China}
\author{Feng Yuan}
\affiliation{Center for Astronomy and Astrophysics and Department of Physics, Fudan University, Shanghai 200438, China}
\author{Xi Lin}
\affiliation{Department of Astronomy, School of Physics, Huazhong University of Science and Technology, Wuhan, 430074, China}
\author{Christian M. Fromm}
\affiliation{Institut f\"ur Theoretische Physik und Astrophysik, Universit\"at W\"urzburg, Emil-Fischer-Str. 31, D-97074 W\"urzburg, Germany}
\affiliation{Institut f\"ur Theoretische Physik, Goethe-Universit\"at Frankfurt, Max-von-Laue-Stra{\ss}e 1, D-60438 Frankfurt am Main, Germany}
\affiliation{Max-Planck-Institut f\"ur Radioastronomie, Auf dem H\"ugel 69, D-53121 Bonn, Germany}
\author{Antonios Nathanail}
\affiliation{Research Center for Astronomy, Academy of Athens, Soranou Efessiou 4, GR-11527 Athens, Greece}
\author{Ziri Younsi}
\affiliation{Mullard Space Science Laboratory, University College London, Holmbury St. Mary, Dorking, Surrey, RH5 6NT, UK}

\correspondingauthor{Yosuke Mizuno, Hong-Xuan Jiang}
\email{mizuno@sjtu.edu.cn, hongxuan\_jiang@sjtu.edu.cn}



\begin{abstract}
Sagittarius~A$^*$, the supermassive black hole at the center of our galaxy, exhibits flares across various wavelengths, yet their origins remain elusive. We performed 3D two-temperature General Relativistic Magnetohydrodynamic (GRMHD) simulations of magnetized accretion flows initialized from multi-loop magnetic field configuration onto a rotating black hole and conducted General Relativistic Radiative Transfer (GRRT) calculations considering contributions from both thermal and non-thermal synchrotron emission processes. Our results indicate that the polarity inversion events from the multi-loop magnetic field configurations can generate $138\,\rm THz$ flares consistent with observations with the help of non-thermal emission.
By tracing the intensity evolution of light rays in GRRT calculations, we identify the precise location of the flaring region and confirm that it originates from a large-scale polarity inversion event.
We observe time delays between different frequencies, with lower-frequency radio flares lagging behind higher frequencies due to plasma self-absorption in the disk. The time delay between near-infrared and 43 GHz flares can reach up to $\sim 50$ min, during which the flaring region gradually shifts outward, becoming visible at lower frequencies. Our study confirms that large-scale polarity inversion in a Standard And Normal Evolution (SANE) accretion flow with a multi-loop initial magnetic configuration can be a potential mechanism driving flares from Sgr~A$^*$.
\end{abstract}



\section{Introduction} \label{sec:intro}
The supermassive black hole (SMBH) at the center of the Milky Way galaxy (i.e., Sagittarius~A$^*$; hereafter, Sgr~A$^*$) is the closest SMBH to Earth. The source has a distance of $D\sim 8\,\rm kpc$ and a mass of $M_{\rm SgrA}\sim 4\times 10^6\,\rm M_\odot$ \citep{EventHorizonTelescopeCollaboration2022}. The emission from the accretion flow around the SMBH forms a bright ring-like structure with a diameter of $51.8\pm2.3\,\rm \mu as$ \citep{EventHorizonTelescopeCollaboration2022}. The proximity of this source makes it a good target for studying the physics of accretion flows in strong gravity. The standard accretion flow model for Sgr~A$^*$ is the radiatively inefficient accretion flow proposed in \citet{Yuan2003}.

The emission from Sgr~A$^*$ is highly variable, which poses challenges in observing it \citep{Collaboration2022c}. Although most of the time, Sgr~A$^*$ stays in a quiescent state, it occasionally shows very strong variability from X-ray to radio bands, so-called flaring state, which can be one order of magnitude higher than that of the quiescent state \citep[e.g.,][]{2001Natur.413...45B,Genzel2003,Schodel2011, Mossoux2020,Abuter2020,Murchikova2021}. Using the very high angular resolution GRAVITY instruments \citep{Abuter2017}, \cite{Abuter2018} detected the orbital motion of hotspots near the last stable orbit of Sgr~A$^*$ during flares in the near-infrared (NIR) band. Polarization observations conducted by ALMA show that the sub-millimeter (mm) emissions from flares are strongly polarized \citep{Wielgus2022}. 
By comparing the light curve between NIR and the radio bands, it appears that the flux variation increases with the observed frequency \citep{Subroweit2017,Abuter2020}. 

Multi-wavelength observations have provided abundant information to help reveal the nature of flares. X-ray and NIR flares usually appear along with each other \citep{2004A&A...427....1E,2006A&A...450..535E}. But the corresponding sub-mm variation has some time delay compared to the X-ray/NIR flares \citep{2003ApJ...586L..29Z, 2006A&A...450..535E, 2006ApJ...644..198Y}. Statistical results by \cite{Witzel2021} suggest the time delay between $230\,\rm GHz$ (sub-mm) and $138\,\rm THz$ (NIR) is $\sim 20$ minutes, while for some particular flares, the delay can be up to $\sim 4.5$ hours. This is consistent with previous observations presented in \citet{Yusef-Zadeh2006}. In this work, light curves at 43 and 22~GHz have been obtained, and it was found that the 43~GHz light curve reaches the peak first. These observations strongly suggest that the flare comes from expanding plasmoids \citep[][]{Yusef-Zadeh2006,Dodds-Eden2010,Rauch2016}.

In the theoretical aspect, various models have been proposed to explain the nature of flares, including shocks in a jet \citep[e.g.,][]{2001A&A...379L..13M,2008A&A...492..337E}, or magnetic reconnection in the accretion flow  \citep[e.g.,][]{2004ApJ...606..894Y,2009MNRAS.395.2183Y,Dodds-Eden2010,2017MNRAS.468.2447P,2014MNRAS.440.2185D,Dexter2020,2020MNRAS.494.5923P,Nathanail2020,Nathanail2022,2021MNRAS.507.5281C,2022ApJ...926..136W, Ripperda2017,Ripperda2021}. While it is believed that magnetic reconnection is a likely physical mechanism for producing the ``hotspot'', this reconnection picture and the production of the hotspot is still not a matter of consensus.

One scenario is that the reconnection occurs in the Magnetically Arrested Disk (MAD) during flux eruption events \citep{Dexter2020,Ripperda2021,Porth2021,2022MNRAS.511.3536S,2024arXiv240410982A,2024A&A...689A.112V, 2025arXiv250107521A}. Saturation of magnetic flux drives eruption events and plasma is expelled from near the horizon, forming a rotating spiral structure. Reconnection occurs via the interface of the magnetically dominated plasma and surrounding fluid.
The polarization study of such vertical flux ropes can robustly produce the loops in the Stokes $\mathcal{Q-U}$ plane \citep{2023arXiv230816740N}, which is one of the observed features during the Sgr~A$^*$ flare \citep{Wielgus2022}. However, in this scenario, the hotspot heated by the reconnection will remain within the accretion flow, which is hard to eject, as suggested by observations. 

In another scenario, reconnection occurs in the coronal region of the accretion flow, where it will further cause the formation of flux ropes. \citet{2009MNRAS.395.2183Y} proposed that turbulence and differential rotation naturally drive magnetic reconnection and flux rope formation in BH accretion flows, even without an initial multi-loop magnetic field. These flux ropes are ejected by magnetic pressure gradients, linking flares to plasmoid ejections. This was later confirmed by 3D simulations \citep{2022ApJ...933...55C}, where flux ropes formed even with a single-loop initial field. \citet{Xi2024} further identified a flux rope whose projected trajectory and super-Keplerian rotation match GRAVITY observations. Recent studies have also explored its radiation and observational signatures \citep[e.g.,][]{Aimar2023,Lin2023,Mellah2023,Xi2024,2024arXiv240714312D}.

Some high-resolution simulations show the possibility of transformation of toroidal magnetic fields into poloidal loops due to effects from mean-field dynamo and magnetorotational instability \citep{2020MNRAS.494.3656L, 2023ApJ...954...40K, 2023ApJ...954L..21G, 2024ApJ...960...97R}.
This enables the creation of the magnetic fields with alternative polarity in the torus and shows "butterfly" diagram\citep{ DelZanna2022, Mattia2020, Mattia2022, 2018ApJ...861...24H, 2024MNRAS.527.3018Z}.
Wind-fed accretion models, like those by \cite{2023MNRAS.521.4277R, 2020ApJ...896L...6R}, usually assume the magnetic field comes from toroidal flux advected by the stellar wind. However, stellar winds might carry magnetic fields with different polarities. If it happens, this could also create a multi-loop structure in the torus, while this is a scenario not yet fully explored in wind-fed accretion research.
Motivated by these, simulations have been performed with a multi-loop structure along the radial direction with an alternating polarity. The accretion of opposite polarity magnetic fields suppresses the accumulation of large magnetic flux and the flow remains in the Standard And Normal Evolution (SANE) state. Consequently, the formation of the jet is also subdued and releases magnetic energy through reconnection \citep{2015MNRAS.446L..61P, Nathanail2020, Nathanail2022, 2023MNRAS.522.2307J, 2019MNRAS.487.4114Y, 2019MNRAS.484.4920Y, 2020MNRAS.494.4203M}. Large-scale polarity inversion events are highly influential on the jet instability and can be generated in the toroidal magnetic field from the poloidal component \citep{2024ApJ...975...57P, 2024MNRAS.532.1522J}. Besides this, polarity inversions have been found to induce frequent magnetic reconnection, leading to the acceleration of non-thermal electrons \citep{2004ApJ...606.1083H, 2014MNRAS.440.2185D, 2016ApJ...826...77B, Ball2018}. This mechanism provides a potential model for Sgr~A$^*$ that differs from the widely discussed flux eruption events in MAD flows.

In our previous work \citep{2023MNRAS.522.2307J, Jiang2024}, we demonstrated that multi-loop magnetic field structures in the torus generate numerous plasmoid chains from polarity inversion events.
We calculated the NIR emission by using general relativistic radiation transfer (GRRT) post-processing of synchrotron radiation with a thermal electron distribution function (eDF) from GRMHD simulations. However, the NIR emissions were too low to compare with the observations of Sgr~A$^*$ flares.  
In this study, we extend our earlier findings by investigating the radiative properties of the accretion flow from polarity inversion events with contributions from both thermal and non-thermal electrons.
We find that this setup can produce powerful NIR flares consistent with observations, with a clear time delay between flares at different frequencies.

The paper is organized as follows: we introduce our GRMHD model and setup for GRRT post-processing calculation in Section~2, present our results in Section~3, and conclude in Section~4.

\section{Methods} \label{sec: method}

\subsection{Numerical setup for GRMHD and GRRT simulations}
To ensure consistency with our previous studies, we adopt a similar setup to perform a 3D GRMHD simulation described in \cite{Jiang2024}. To avoid the strong magnetic dissipation seen in the small loop length cases and the powerful jet associated with the large loop length cases, we focus on a multiple magnetic loop configuration with alternating polarities with moderately long wavelength, which is given by \citep{Nathanail2020}:
%
\begin{equation}
\begin{aligned}
    A_{\rm \phi}\propto& (\rho - 0.01)(r/r_{\rm in})^3\sin^3\theta \exp{(-r/400)}\\
    &\cos((N-1)\theta)\sin(2\pi(r-r_{\rm in})/\lambda_{\rm r}),
\end{aligned}
\end{equation}
where $\lambda_{\rm r}$ sets the size of the magnetic loops. We choose $\lambda_{\rm r}=30\,r_{\rm g}$ in this work, where $r_{\rm g}\equiv GM/c^2$ is the gravitational radius, and where $G$ is the gravitational constant, $M$ the mass of the BH, and $c$ the speed of light. The value of $A_{\rm \phi}$ is determined by setting the minimum of plasma $\beta_{\rm min}=100$, where $\beta \equiv p_{\rm g}/p_{\rm mag}$.
We use Fishbone-Moncrief hydrostatic equilibrium torus \citep{1976ApJ...207..962F}. The torus parameters are set to be $r_{\rm in} = 20\, r_{\rm g}$ and $r_{\rm max} = 40\, r_{\rm g}$. The spin of the BH is set to be $a=0.9375$.
The details of the GRMHD code setup and numerical resolution are presented in Appendix~\ref{Sec: kharma} and \ref{sec:res_check}. Two-temperature treatment for the electron temperature is included following \cite{Ressler2015}. The details of this treatment are described in Sec.~\ref{sec:2T}.

To study the emission properties from the accretion flow, GRRT post-processing is implemented with the \texttt{BHOSS} code \citep{2012A&A...545A..13Y,2020IAUS..342....9Y}. In GRRT calculations, the radiative transfer coefficients are determined following \cite{Marszewski2021}. We use a $\kappa$ eDF to incorporate non-thermal emission in the calculation, which has an acceptable range within $3.5\leq\kappa\leq7.5$. The value of $\kappa$ is calculated using a subgrid model based on turbulent plasmas (see Sec.~\ref{sec:2T} for details). The values of $\kappa$ may be outside this range for magnetization $\sigma (=b^2/\rho)>5$ or $\sigma\ll1$. The non-thermal synchrotron emissions $(j_{\rm kappa})$ from these regions are always set to zero (see more discussion on the choice of the range of magnetization in Appendix~\ref{sec: sigma_cut}). In contrast, the thermal synchrotron emission ($j_{\rm Th}$) is set to zero only for highly magnetized regions with $\sigma>5$. 
Accordingly, the total emissivity is calculated as $j_{\rm tot}=(1-\tilde{\epsilon})j_{\rm Th}+\tilde{\epsilon}j_{\rm kappa}$, where $\tilde{\epsilon}$ is the fraction of energy contributed by magnetic field (for a detailed definition see Sec.~\ref{sec:2T} and \cite{Fromm2022}). 
For the high frequency emission a cooling effect is added by following \cite{2022MNRAS.511.3536S} with details described in Appendix~\ref{sec: cooling}.
We consider Sgr~A$^*$ as a target source where the BH mass is $M_\bullet=4.14\times 10^6\,\rm M_{\odot}$, distance is $D_{\rm SgrA}=8.127\,\rm kpc$ \citep{Abuter2019_BHMASS}. 
We use the field of view of $30\,r_{\rm g}\times 30\,r_{\rm g}$ ($150\,\rm \mu as \times 150\,\rm \mu as$) with a resolution of $1024\times 1024$. The inclination angle between the line of sight and the BH spin direction in the GRRT calculation is set to be $25^\circ$.
Additionally, the mass unit for scaling of the GRMHD simulations is obtained by fitting with a flux of $3.5\,\rm Jy$ at 230~GHz.
From this, we roughly fit the spectral energy distribution (SED) from millimeter frequencies to the NIR band (see Appendix~\ref{sec: sed}).
GRRT post-processing uses the GRMHD data from $8,000-11,000\,GM/c^3$, during which more polarity inversion events and stronger NIR flares are seen.

\subsection{Turbulence and reconnection models of thermal and non-thermal eDF}
\label{sec:2T}

We obtain electron temperature directly by solving the electron entropy equation with the turbulent and reconnection heating prescriptions following \cite{2019PNAS..116..771K, Rowan2017}. The detailed introduction of these two prescriptions for electron heating see \cite{2023MNRAS.518..405D}.

For the study of the NIR flares from Sgr~A$^*$, non-thermal emission plays a dominant role \citep{Yuan2003, 2022MNRAS.511.3536S}. Here we introduce our subgrid models for the non-thermal emission, based on Particle-In-Cell (PIC) simulations of magnetic reconnection and turbulence in plasma.
Accordingly, in this work, we use kappa eDF for the calculation of synchrotron radiation by electrons. The kappa eDF is a hybrid eDF model with the combination of Maxwell-J\"uttner and power-law eDFs \citep{2006PPCF...48..203X, Davelaar2019, Fromm2022}.
In kappa eDF, $\kappa$ displays analogous behavior with the slope $p=\kappa-1$ in power-law eDF. The value of $\kappa$ reflects the microphysics of the local plasma environment. In this work, we consider a sub-grid model obtained from Particle-in-Cell (PIC) simulations of turbulent plasma \citep{Meringolo2023}. 

The explicit expression of kappa eDF is written as follows:
\begin{equation}
    \frac{dn_{\rm e}}{d\gamma_{\rm e}}=\frac{N}{4\pi}\gamma_{\rm e}\sqrt{\gamma_{\rm e}^2-1}\left(1+\frac{\gamma_{\rm e}-1}{\kappa w}\right)^{-(\kappa+1)},
\end{equation}
where $n_{\rm e}$ is the electron number density, $\gamma_{\rm e}$ is the electron Lorentz factor, $w$ is the width
of the kappa distribution, and $N$ is a normalization factor (see \cite{Pandya2016} for detail). $\kappa$ displays analogous behavior with the power-law index $p$ in power-law eDF. For $\kappa\gg10$, kappa eDF decays to Maxwell-J\"uttner one (see the Fig.~4 in \cite{Fromm2022}). The energy width of the kappa eDF is set by the value $w$ \citep{Fromm2022}, which corresponds to the dimensionless electron temperature $\Theta_{\rm e}$ when $\kappa$ is infinity. Finally, the non-thermal emissivity of the non-thermal electrons is calculated following \cite{Marszewski2021}. 

We adopt two functional forms of $\kappa$, considering fitting formula from PIC simulations of special-relativistic turbulence \citep[][$\kappa_{\rm tur}$]{Meringolo2023} and magnetic reconnection \citep[][$\kappa_{\rm rec}$]{Ball2018}. They are given by:
\begin{equation}
    \kappa_{\rm tur}(\beta,\sigma)=2.8+\frac{0.2}{\sqrt{\sigma}} + 1.6\sigma^{-6/10}\tanh{\left(2.25\beta\sigma^{1/3}\right)}, \label{Eq: kappa_tur}
\end{equation}
\begin{equation}
    \kappa_{\rm rec}(\beta,\sigma)=2.8+\frac{0.7}{\sqrt{\sigma}} + 3.7\sigma^{-0.19}\tanh{\left(23.4\beta\sigma^{0.26}\right)},
    \label{Eq: kappa_rec}
\end{equation}

where $\sigma$ and $\beta$ are the magnetization and plasma beta.

Based on \cite{Davelaar2019, 2018A&A...612A..34D}, we use the non-thermal energy efficiency to determine the width $w$ of the kappa distribution. The original form in \cite{Davelaar2019} is written as
\begin{equation}
    w=\frac{\kappa-3}{\kappa}\Theta_{\rm e}+\tilde{\epsilon}\frac{\kappa-3}{6\kappa}\frac{m_{\rm p}}{m_{\rm e}}\sigma, \label{Eq: w}
\end{equation}
where the first and second terms correspond to the thermal and non-thermal energy terms, respectively; $m_{\rm p}$ and $m_{\rm e}$ are the proton and electron mass. Here, we assume that the non-thermal energy comes from the magnetic energy. The fitting formula of emissivity in kappa eDF requires $3.5\leq\kappa\leq7.5$ \citep{Marszewski2021}. In both cases, we constrain the value of $\kappa$ within this range. $\tilde{\epsilon}$ represents the fraction of magnetic energy attributed to the width of the kappa distribution \citep{Fromm2022}. Specifically, it accounts for the proportion of magnetic energy relative to the total energy in the system. In previous works, $\tilde{\epsilon}$ is usually set with constant values (0, 0.5, 1.0) \citep{Fromm2022, Cruz-Osorio2022}. However, the non-thermal acceleration efficiency usually varies across different plasma parameters. \cite{Meringolo2023} and \cite{Ball2018} provided fitting formulas for the non-thermal energy efficiency of turbulence and reconnection models:
\begin{equation}
    \tilde{\epsilon}_{\rm tur} = 1.0 - \frac{0.23}{\sqrt{\sigma}} + 0.5 \sigma^{1/10}\tanh{\left[-10.18\beta \sigma^{1/10}\right]}, \label{Eq: epsilon_tur}
\end{equation}
\begin{equation}
    \tilde{\epsilon}_{\rm rec} = 1.0 - \frac{1}{4.2\sigma^{0.55}+1} + 0.64 \sigma^{0.07}\tanh{\left[-68\beta \sigma^{0.13}\right]}, \label{Eq: epsilon_rec}
\end{equation}
We assume that the energy of the non-thermal electrons comes from the magnetic energy. Therefore, in Eq.~\ref{Eq: w} we use this variable $\tilde{\epsilon}$ to adjust the contribution of non-thermal electrons locally. 

Following \cite{2022A&A...660A.107F}, the total emissivity $j_{\rm tot}$ and absorptivity $\alpha_{\rm tot}$ in this work include two components, which are calculated based on thermal and kappa eDF. They are calculated with:
\begin{equation}
\begin{aligned}
    j_{\rm tot}&=(1-\tilde{\epsilon})j_{\rm Th}+\tilde{\epsilon}j_{\rm kappa},\\
    \alpha_{\rm tot}&=(1-\tilde{\epsilon})\alpha_{\rm Th}+\tilde{\epsilon}\alpha_{\rm kappa},
\end{aligned}
\label{Eq: emi&alpha}
\end{equation}
where $\tilde{\epsilon}$ uses the same variable ones in Eq.~\ref{Eq: epsilon_tur} and Eq.~\ref{Eq: epsilon_rec}. The emissivities and absorptivities from both the thermal and non-thermal electrons are combined before being used in the GRRT calculations.

In the relatively weaker magnetized regions ($\sigma < 0.01$), we only use emissivity and absorption from thermal eDF, while the hybrid method in Eq.~\ref{Eq: emi&alpha} is applied in the strongly magnetized regions.

Our subgrid models are implemented using criteria based on the local, macroscopical state of the plasma. Non-thermal particles are injected when large-scale fluid variables indicate conditions are favorable for reconnection, as the simulation itself does not resolve the microphysical structures (e.g., current sheets) where this acceleration occurs. This approach, therefore, uses a parameterization of the underlying findings of earlier small-scale studies (as discussed above), and a more self-consistent treatment of these cross-scale interactions remains a critical challenge for the required numerical resolution.

\section{Results}\label{sec: results}
\subsection{Multi-frequency light curves}

\begin{figure*}
    \centering
    \includegraphics[width=.9\linewidth]{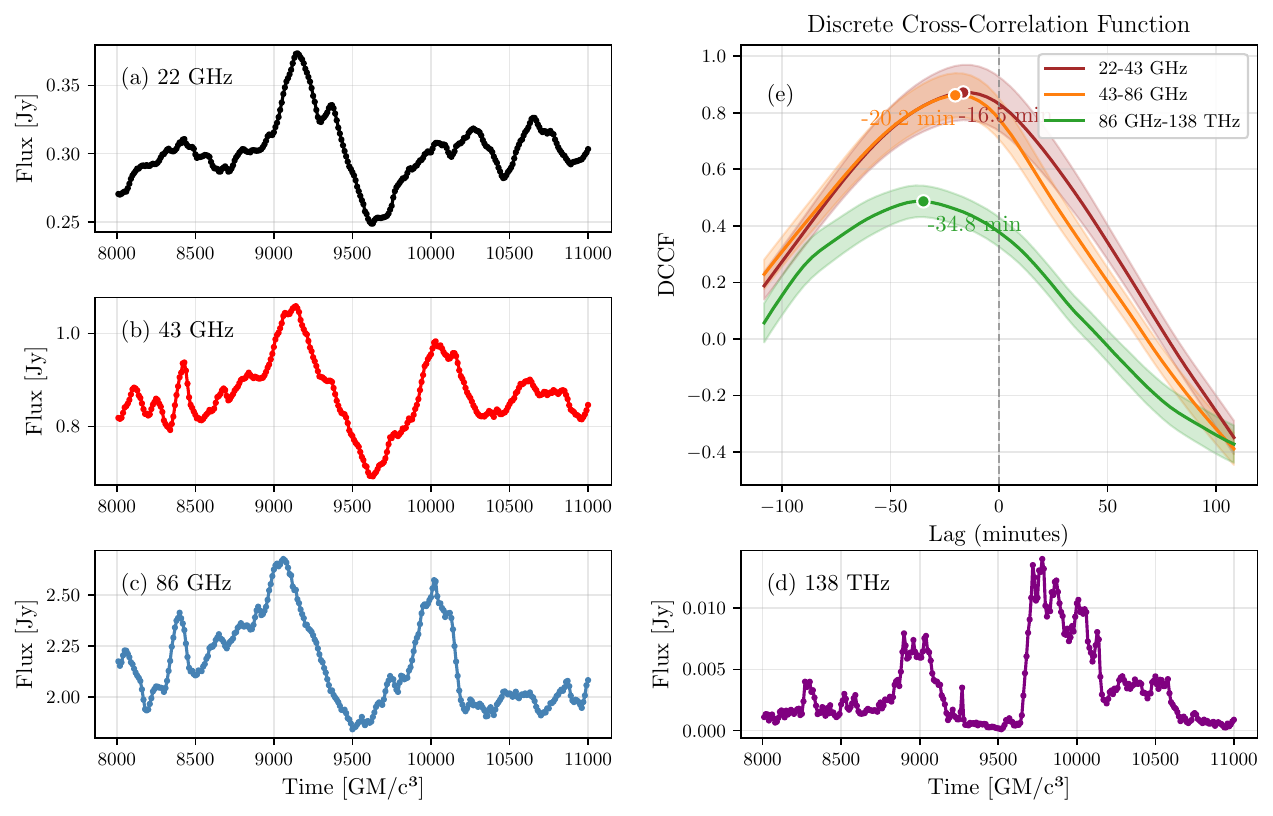}
    \caption{Multi-frequency light curves and Discrete Cross-Correlation Function (DCCF). Panels (a)-(d) show light curves from GRRT post-processing calculations at $22\,\rm GHz$ (black), $43\,\rm GHz$ (red), $86\,\rm GHz$ (blue), and $138\,\rm THz$ (NIR, purple). Panel (e) presents the DCCFs for $22$-$43,\rm GHz$ (blue), $43$-$86\,\rm GHz$ (orange), and $86,\rm GHz$-$138\,\rm THz$ (green), with the shaded regions indicating the $1\sigma$ uncertainty range.}
    \label{fig: f1}
\end{figure*}

In previous works, it has been demonstrated that the accretion flows from a multi-loop magnetic configuration trigger frequent magnetic reconnection through polarity inversion \citep[e.g., ][]{Nathanail2020, 2014MNRAS.440.2185D}. In polarity inversion events, numerous plasmoid chains and current sheets are generated, leading to flaring events consistent with those observed in Sgr~A$^*$ \citep[e.g., ][]{Abuter2020}. During this process, non-thermal electrons are accelerated in regions associated with polarity inversions, injecting magnetic energy into the non-thermal components of the eDF and driving the luminous flares seen in the NIR band. 

By applying GRRT post-processing to non-thermal electrons, we obtain multi-frequency light curves spanning the simulation period with the highest occurrence of polarity inversions (from $8,000$ to $11,000\,GM/c^3$), as shown in Fig.~\ref{fig: f1}. From panels (a) to (d), the light curves at frequencies of 22~GHz, 43~GHz, 86~GHz, and 138~THz (NIR) are presented in black, red, blue, and purple lines, respectively. From the $138\,\rm THz$ light curve, we identify three major flaring events at $t \sim 8320$, $9000$, and $9780\,GM/c^3$. The peak flux at $138\,\rm THz$ reaches up to $\sim14 \,\rm mJy$, while the minimum flux drops below $0.1\,\rm mJy$. The evolution of the light curve of the $138\,\rm THz$ exhibits good agreement with near-infrared observations of Sgr~A$^*$ \citep{Abuter2020}. 

As illustrated in panels (a)-(d) of Fig.~\ref{fig: f1}, a comparison of the millimeter radio and NIR light curves reveals a frequency-dependent time-lag between them.
Following previous works \citep[e.g., ][]{Witzel2021, 2024arXiv240410982A}, we use the cross-correlation function (CCF) to quantitatively measure the time delay between the light curves of different frequencies (see Appendix~\ref{Sec: CCF} and \ref{sec: statistics} for more detail).
In panel (e) of Fig.~\ref{fig: f1}, we present the discrete cross-correlation functions (DCCFs) between the light curves at 22~GHz, 43~GHz, 86~GHz, and 138~THz, where the shaded region represents the error in each DCCF bin.
The DCCFs shown in panel (e) quantify the delays, yielding measured lags of $\sim16.5$ minutes between $22$ and $43$~GHz, $\sim20.2$ minutes between $43$ and $86$~GHz, and $\sim34.8$ minutes between $86$~GHz and $138$~THz, demonstrating longer delays at lower frequencies. This trend is particularly evident during the flaring event around $\sim 9,000\,\rm GM/c^3$, where the lower-frequency radio emission peaks later than the higher-frequency NIR component. It is consistent with observations showing that flares at lower frequencies typically lag behind those at higher frequencies \citep[e.g.,][]{Yusef-Zadeh2006,2008ApJ...682..361Y}. 

\begin{figure*}
\centering
 	\includegraphics[width=.49\linewidth]{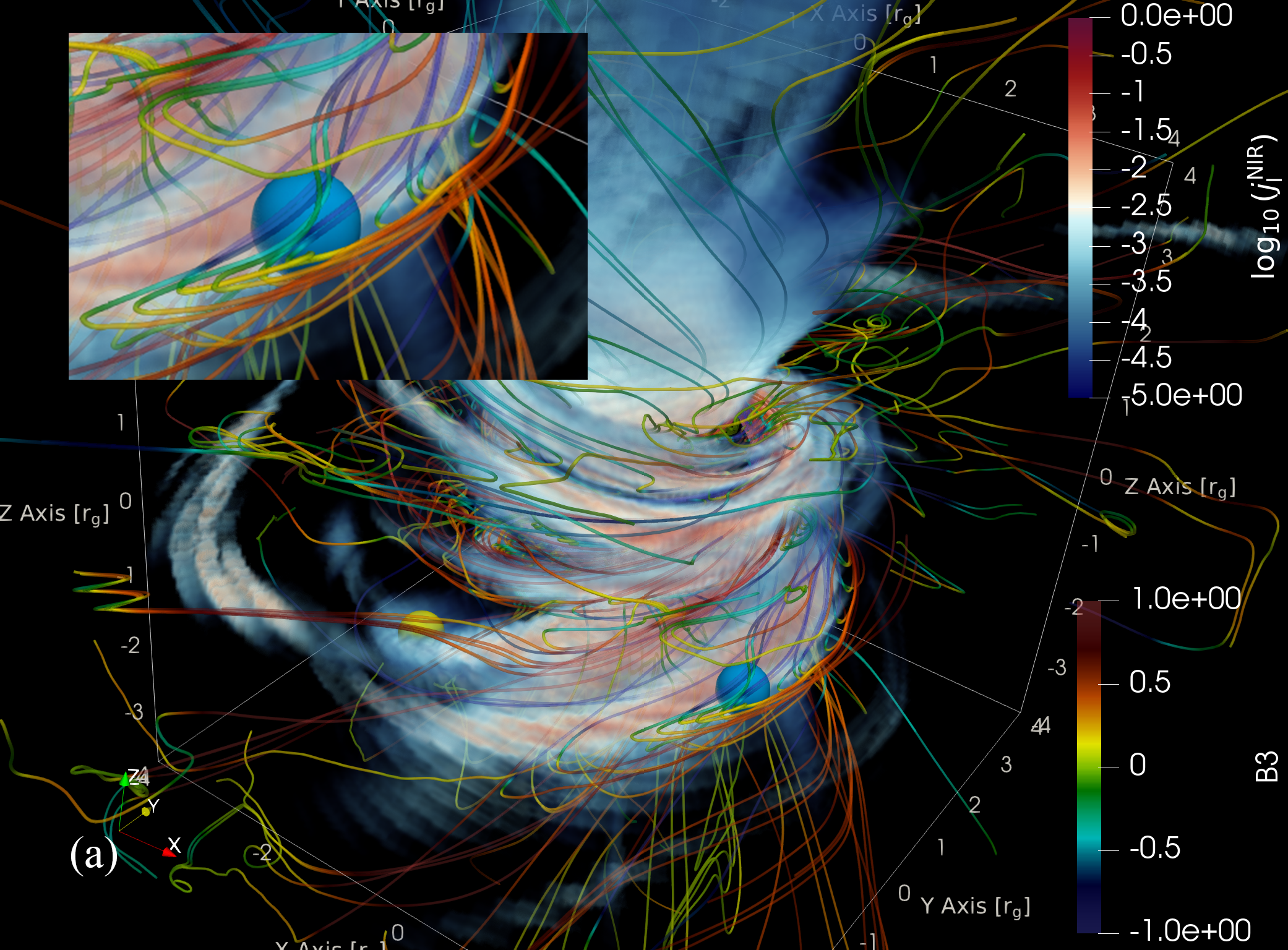}
 	\includegraphics[width=.49\linewidth]{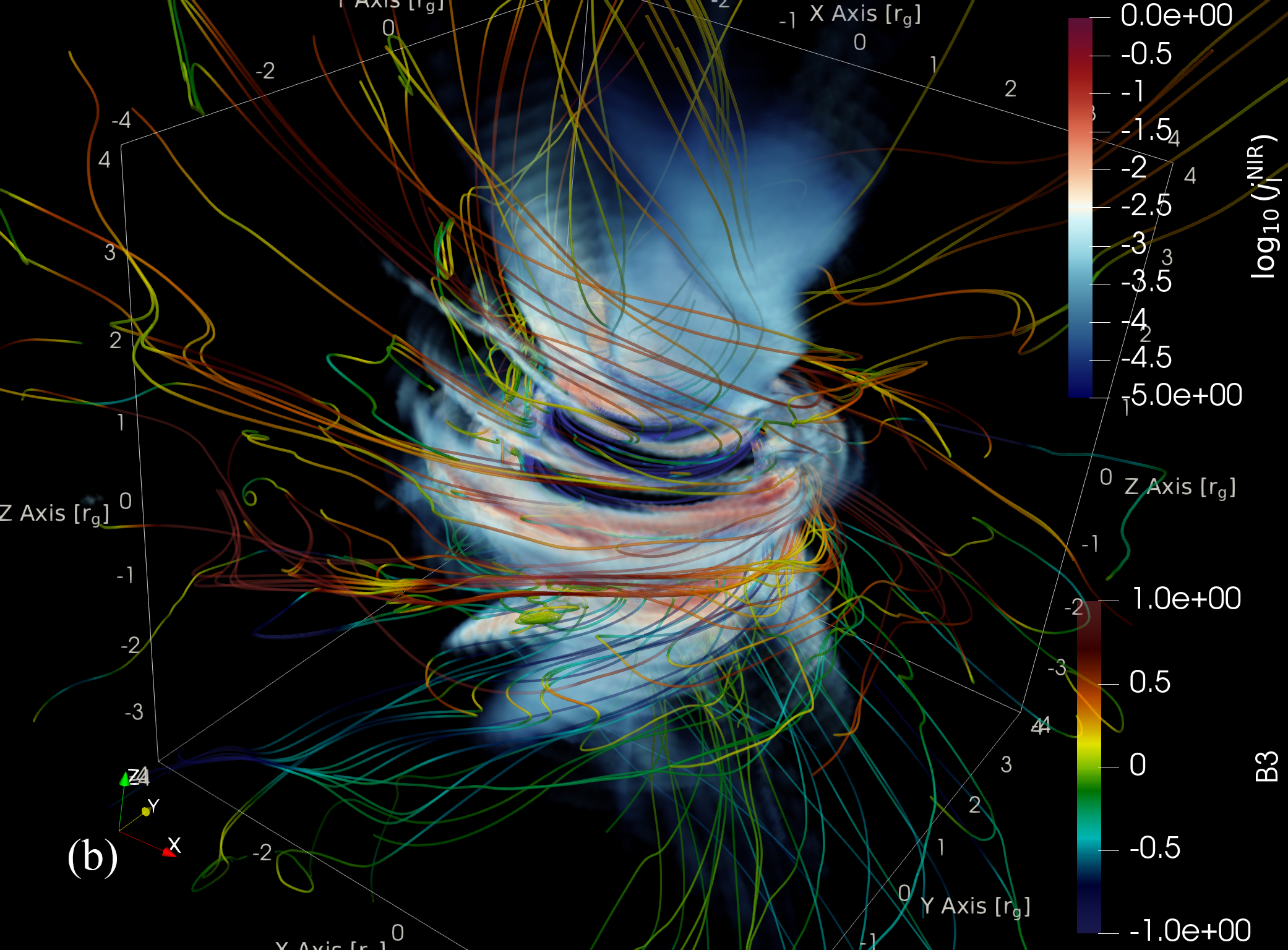}
 	\includegraphics[height=.4\linewidth]{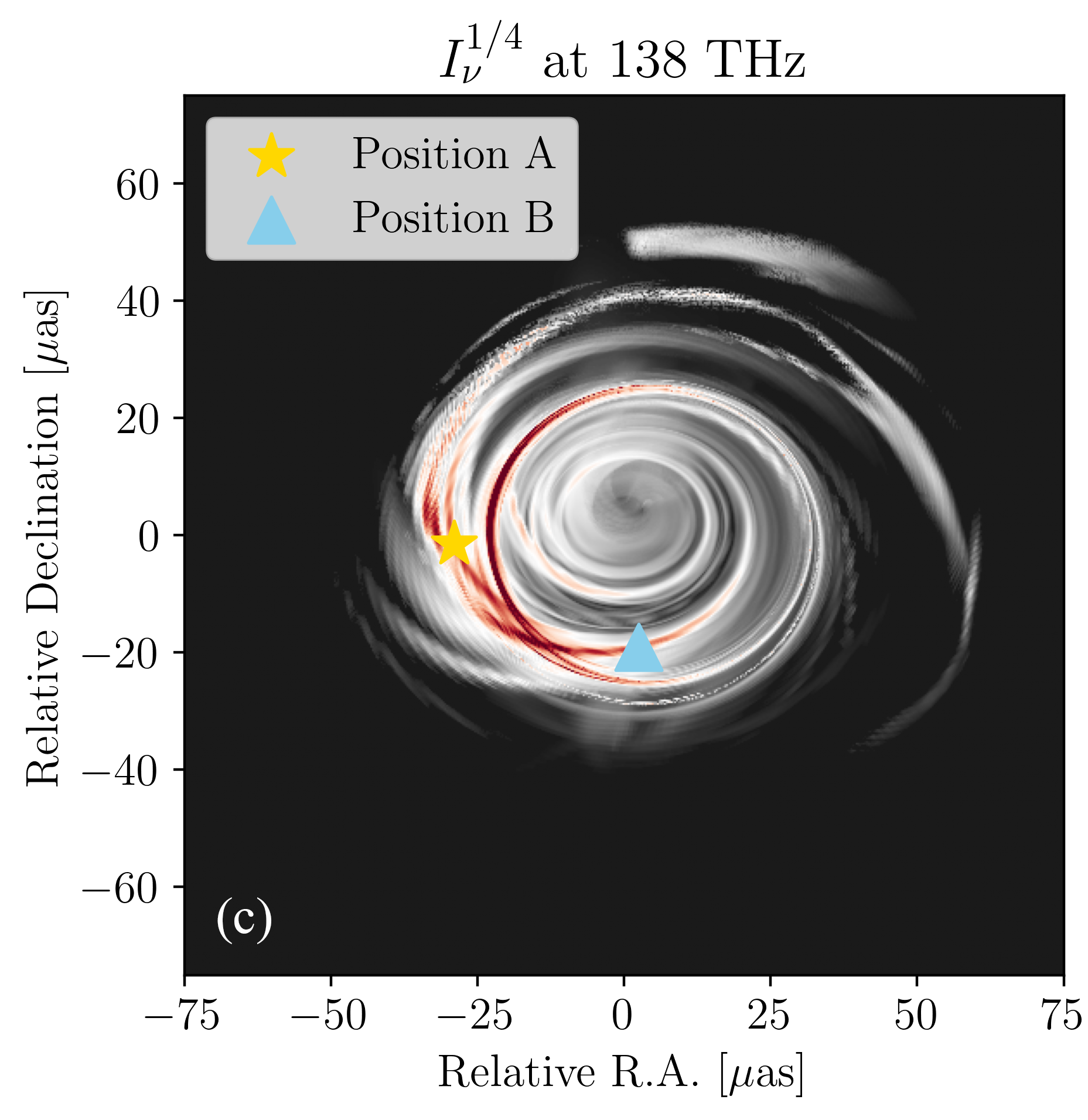}
 	\includegraphics[height=.4\linewidth]{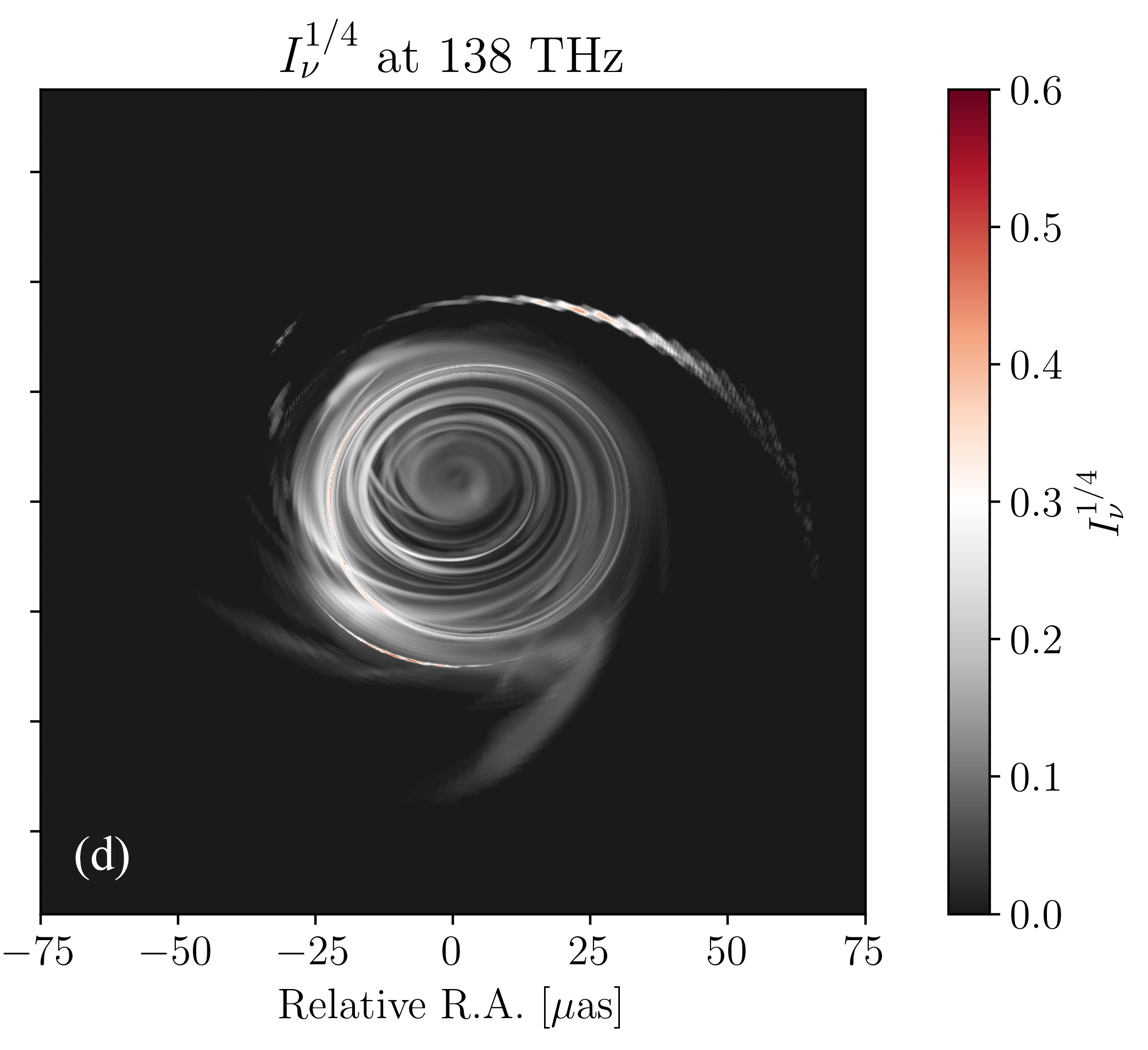}
 	\caption{Large-scale polarity inversion as a source of observed NIR flaring event. Panels (a) and (b) show the normalized NIR emissivity ($j_{\rm I}^{\rm NIR}$, normalized by its maximum value) and the corresponding magnetic field lines during the flaring (left) and quiescent states (right) at $t = 8,320\,\rm M$ and $9,500\,\rm M$, respectively. The color of the magnetic field lines represents the relative strength of the toroidal component. To clearly illustrate the polarity inversion and the origin of the NIR flare, we zoom in to the flaring region, where the toroidal magnetic field lines twist and reverse polarity in the upper left corner of Panel (a). This extended polarity inversion is not clearly visible in the quiescent state shown in Panel (b). The intensity maps $I_{\nu}^{1/4}$ from GRRT calculations at flaring and quiescent states are presented in Panels (c) and (d), where $I_{\nu}^{1/4}$ is normalized to the maximum value in Panel (c). In panel (c), two positions, labeled as Position A and Position B, are marked by a yellow triangle and a blue star, respectively. Their corresponding emitting regions are shown in panel (a) as yellow and blue spheres in the volume rendering image.}
 	\label{Fig: f2}
\end{figure*}

\begin{figure}
    \centering
    \includegraphics[width=\linewidth]{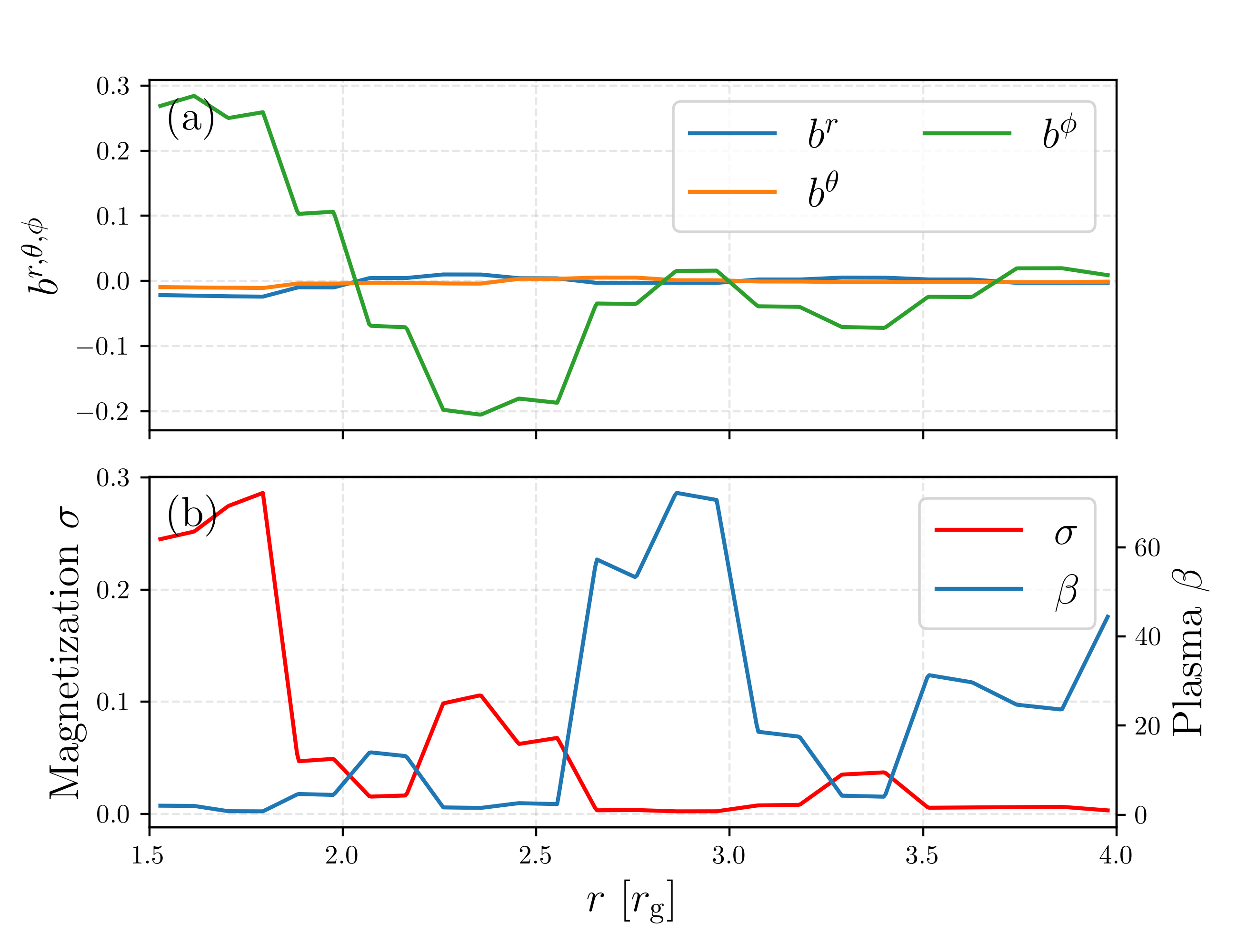}
    \caption{One-dimensional (1D) radial profile taken from the zoomed-in region shown in Figure~\ref{Fig: f2}(a). This figure is plotted with close $\theta$ and $\phi$ angles with position B in Figure~\ref{Fig: f2}(a). Panel (a) displays the components of the magnetic field, clearly illustrating the reversal of the toroidal field's direction during a large-scale polarity inversion event. Panel (b) plots the corresponding radial profiles for the plasma magnetization ($\sigma$, red) and the plasma $\beta$ (blue).}
    \label{Fig: 1D_cut}
\end{figure}

\begin{figure*}
\centering

\includegraphics[width=.49\linewidth]{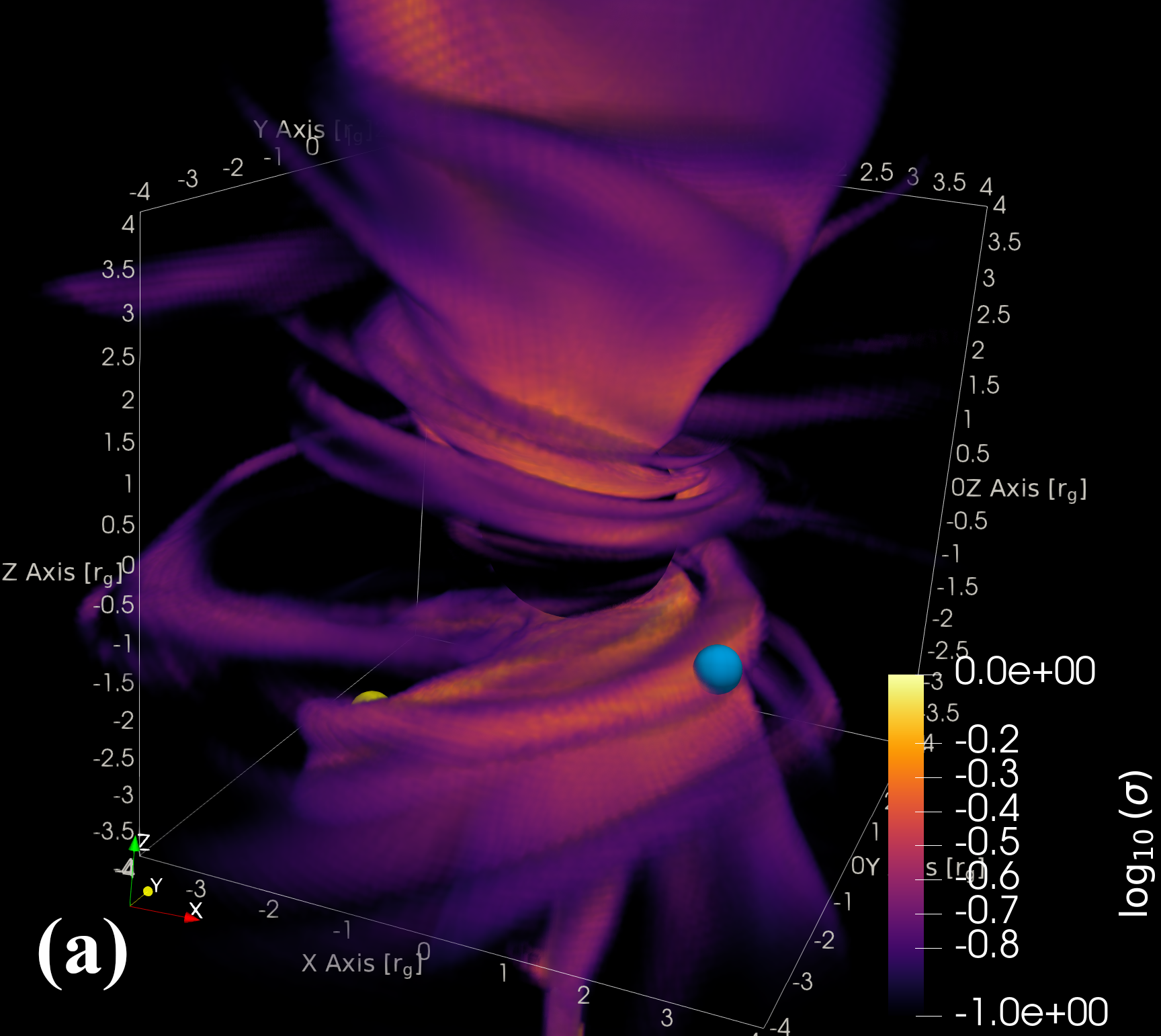}
\includegraphics[width=.49\linewidth]{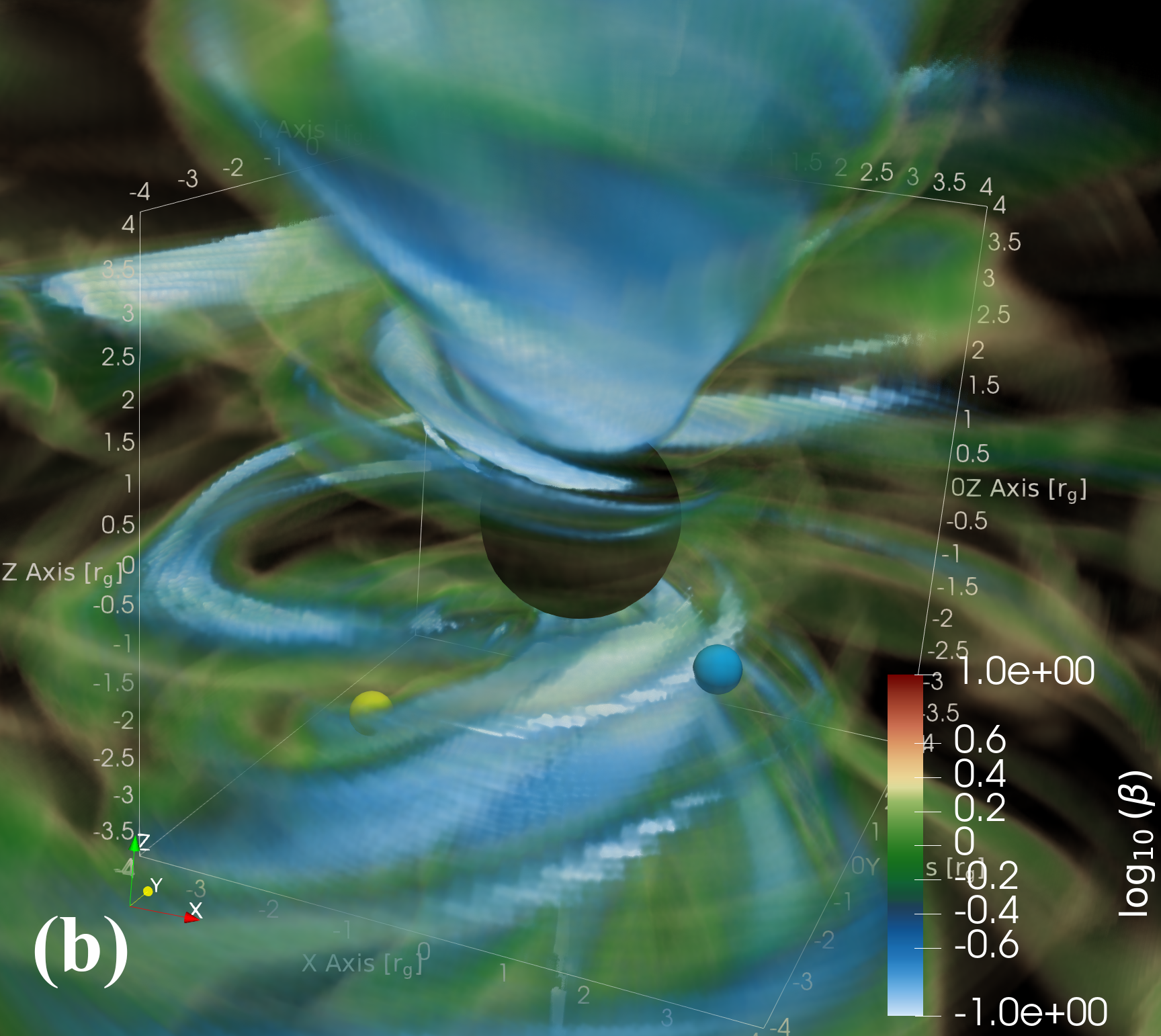}
\includegraphics[width=.49\linewidth]{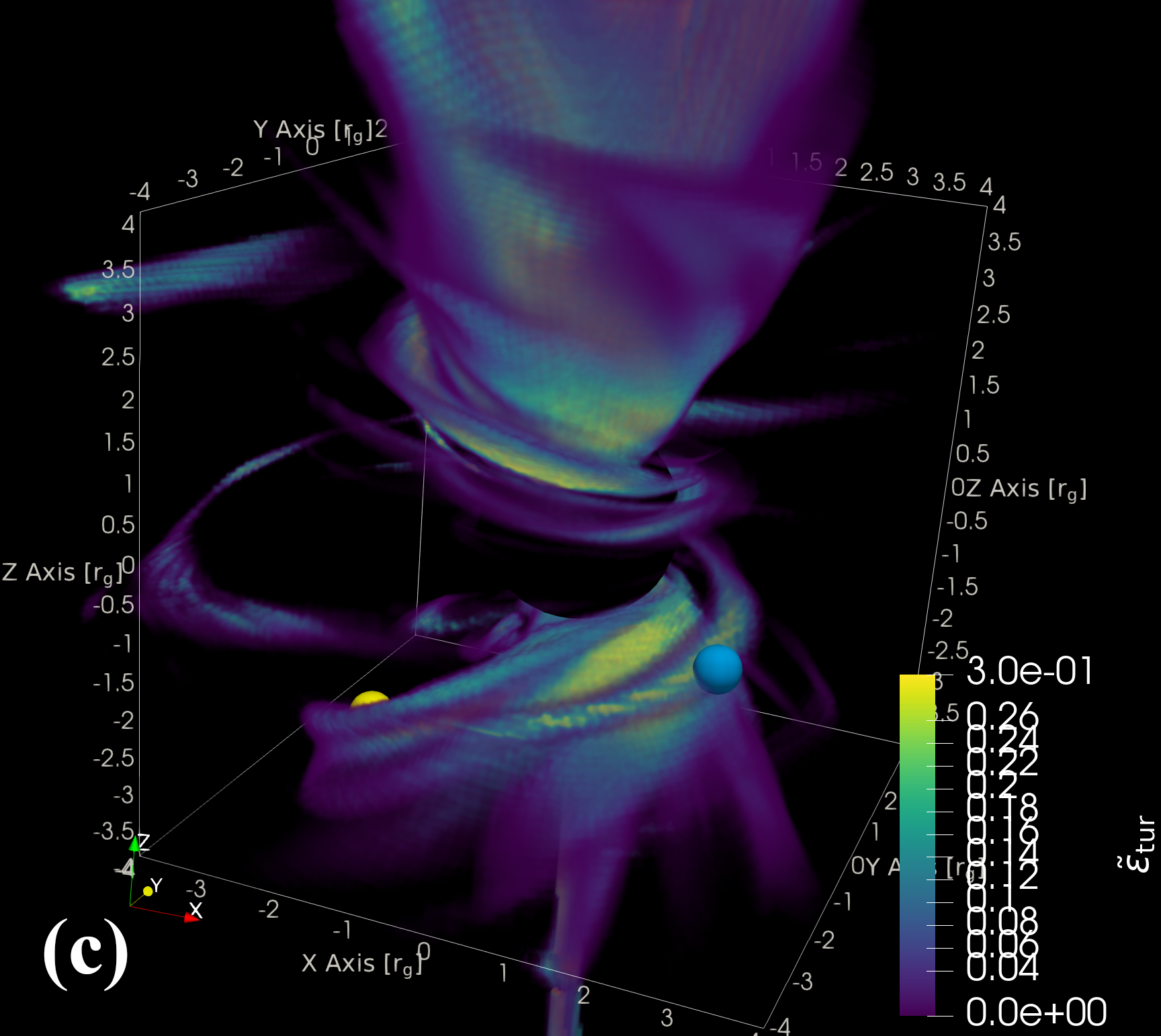}
\includegraphics[width=.49\linewidth]{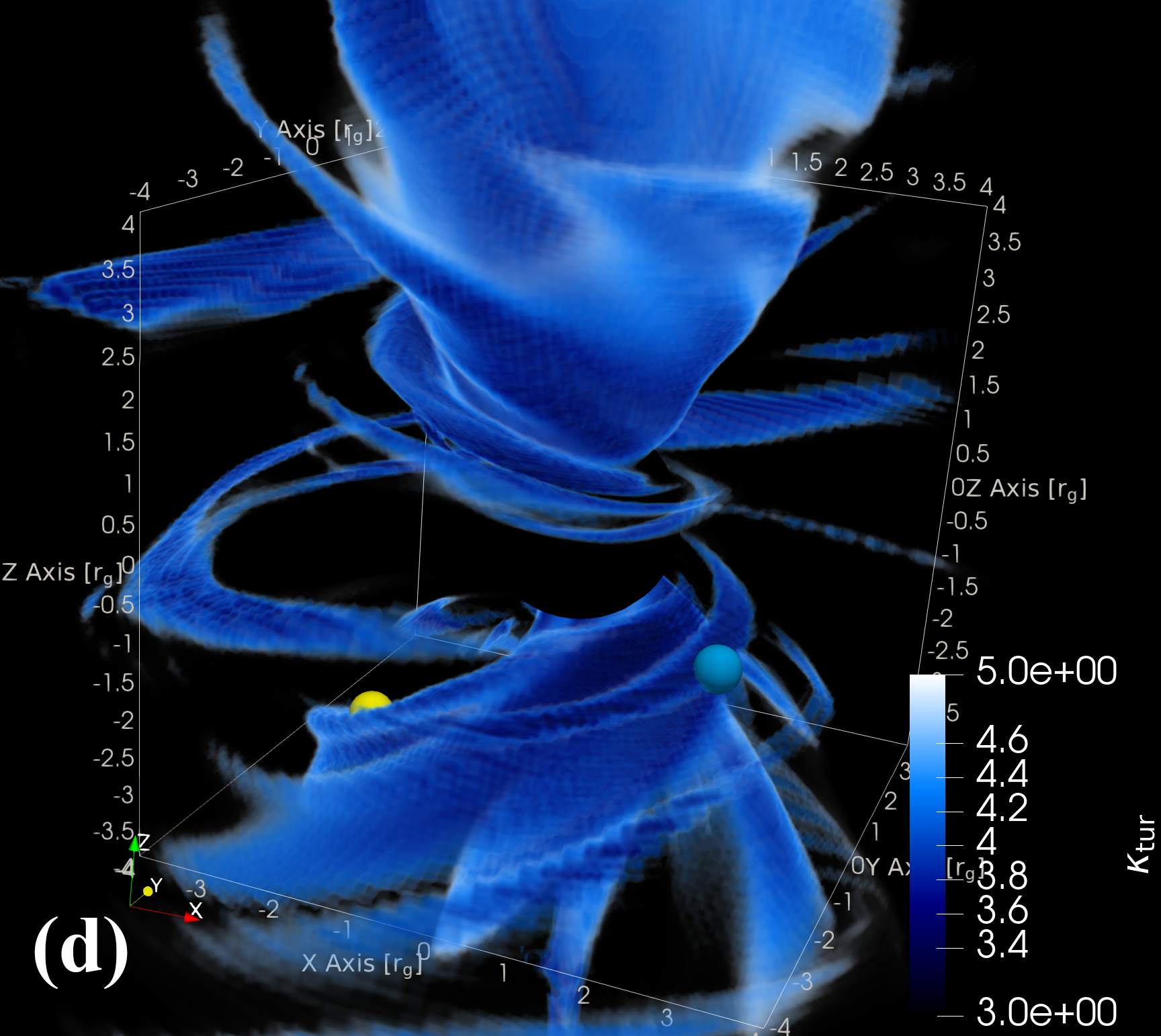}
 	\caption{Plasma properties and non-thermal electron signatures in the reconnection site.
 Panel (a) shows the plasma magnetization, $\sigma$; 
(b) shows the plasma beta, $\beta$; (c) shows the non-thermal acceleration efficiency from our turbulence model; and (d) shows the resulting non-thermal index, $\kappa$. The yellow and blue spheres are identical to those in Fig.~\ref{fig: f1}(b) and are included to mark the location of the polarity inversion event.}
 	\label{Fig: beta&sigma}
\end{figure*}

\begin{figure*}
    \centering
    \includegraphics[height=.35\linewidth]{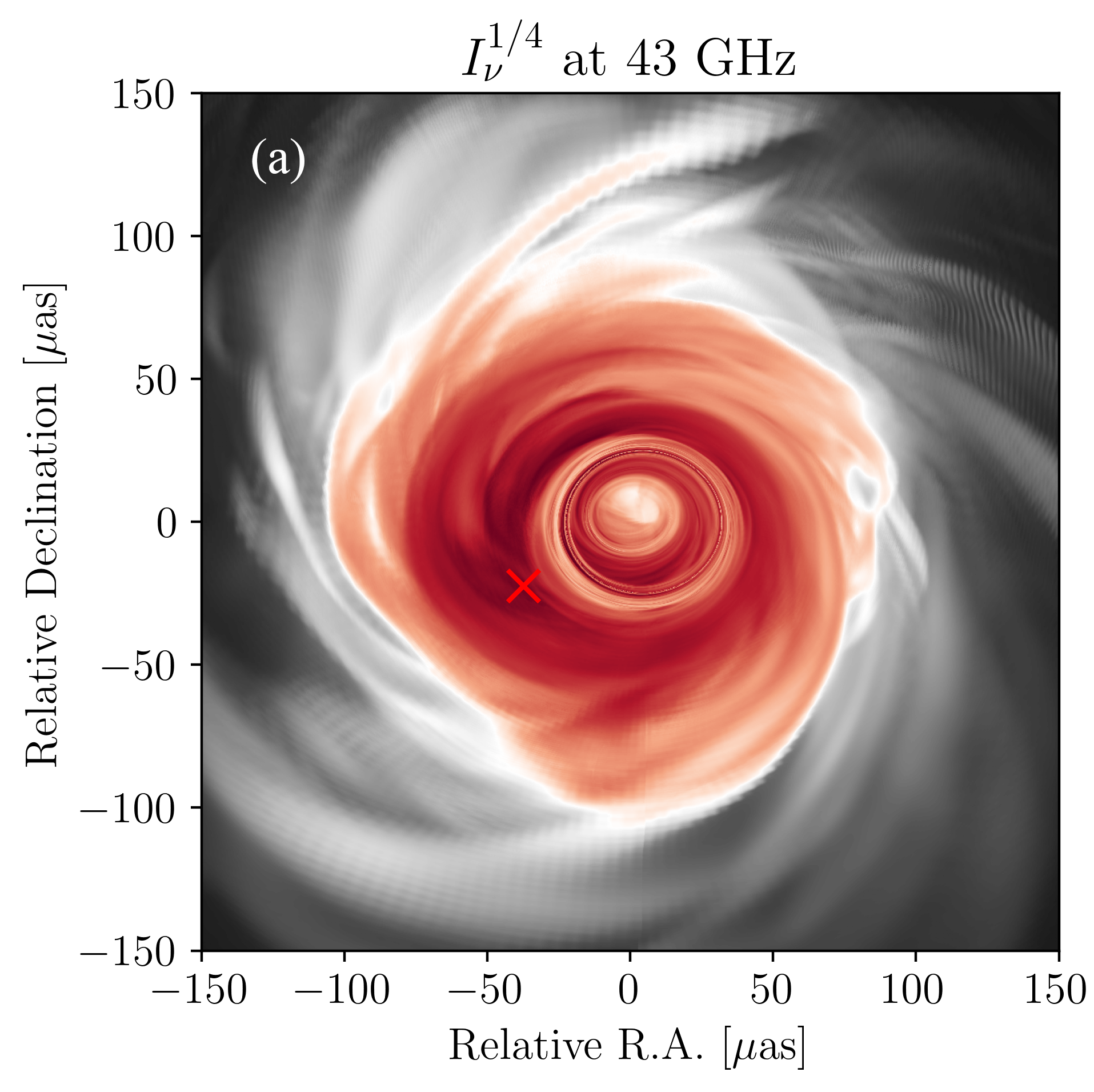}
    \includegraphics[height=.35\linewidth]{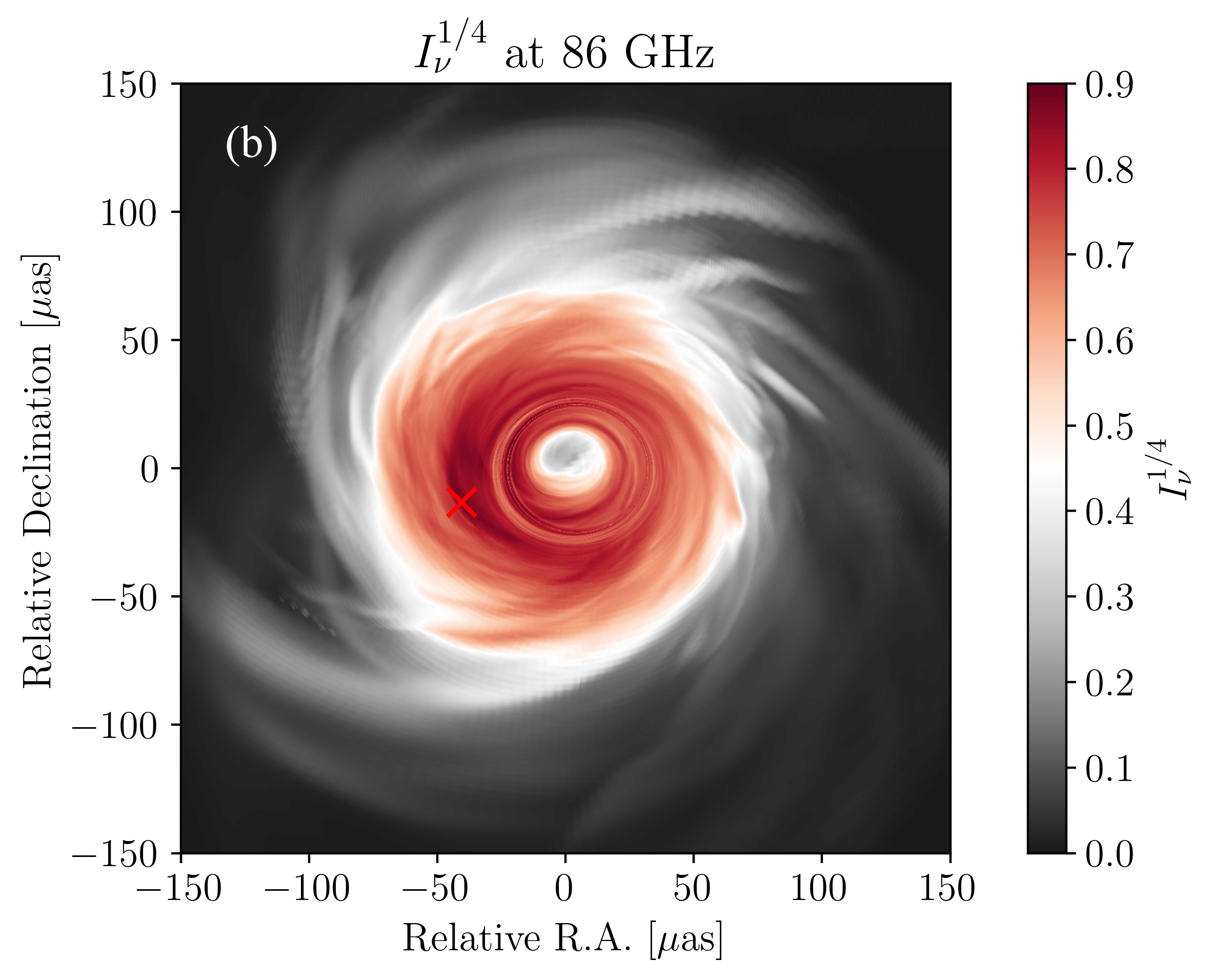}
    \includegraphics[width=.8\linewidth]{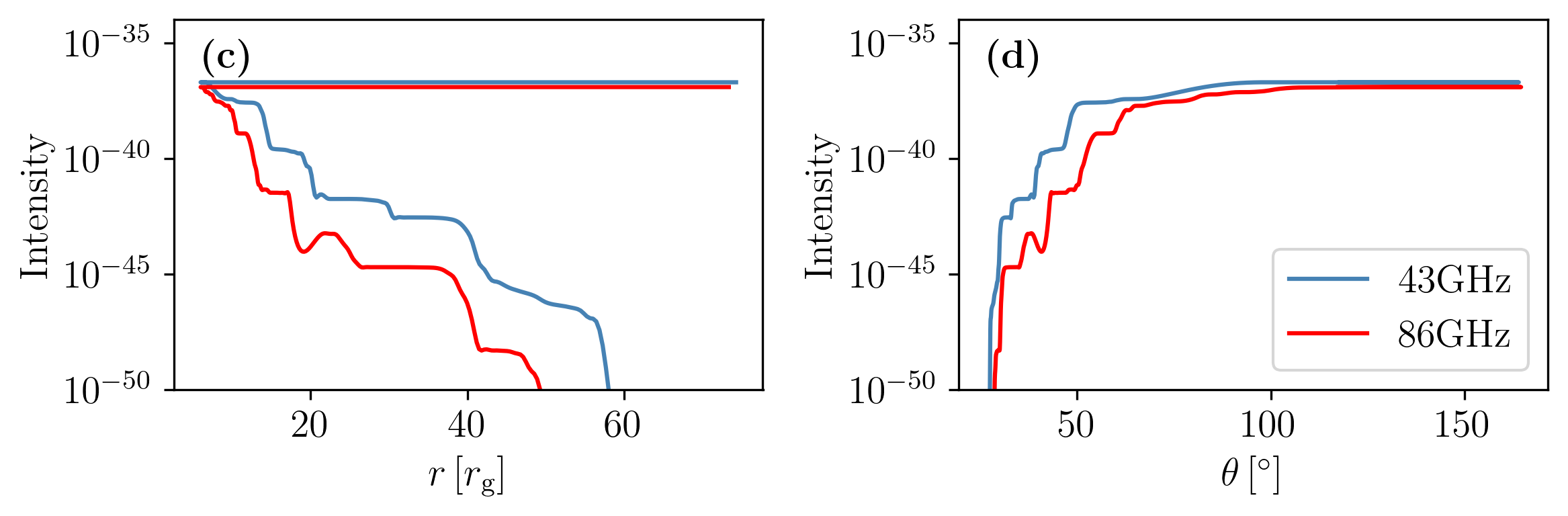}
    \caption{Outward shift in the location of the peak emission in radiative transfer. Panels (a) and (b) show the GRRT images at $t = 8,380$ and $8,420 \,GM/c^3$, corresponding to the 43 and $86\,\rm GHz$ flares, respectively. The brightest pixel in each panel is marked with a red cross. Panels (c) and (d) illustrate the intensity evolution along the light-ray paths of the marked pixels. The blue and red curves represent the intensity tracks of the 43 and $86\,\rm GHz$ flares, respectively. Panel (c) shows the radial evolution, while panel (d) depicts the evolution in the polar direction.}
    \label{fig: f3}
\end{figure*}

\begin{figure}
    \centering
    \includegraphics[width=\linewidth]{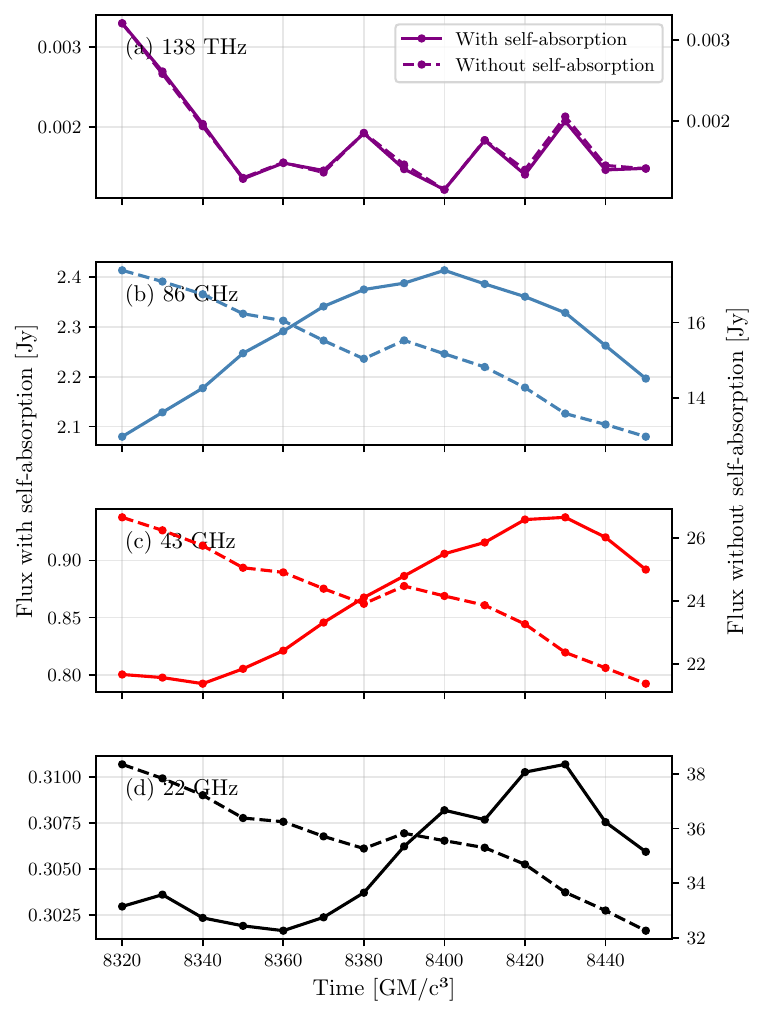}
    \caption{Common origin of the multi-frequency flares and the self-absorption induced time delay. Panels (a) to (d) correspond to light curves at $138\,\rm THz$, $86\,\rm GHz$, $43\,\rm GHz$, and $22\,\rm GHz$, shown in purple, blue, red, and black, respectively. The light curves are obtained from GRRT post-processing calculations, including (solid lines) and excluding (dashed lines) the effects of plasma self-absorption. 
    }
    \label{fig: f4}
\end{figure}

\subsection{Origin of NIR flares}\label{Sec: NIR_origin}

In GRMHD simulations with multiple magnetic loops, polarity inversion events are found to possess strong and ordered magnetic fields, as well as relatively high electron temperatures and densities in the jet sheath region, which may be related to the flaring activities in Sgr~A$^*$ \citep{2015MNRAS.446L..61P, Nathanail2020}. 

From panel (d) of Fig.~\ref{fig: f1}, our simulation shows a NIR light curve exhibiting a flux increase of nearly two orders of magnitude during flares compared to the quiescent state. The amplification, absolute flux during flare, and flare duration are quite close to those reported in \cite{Abuter2020}. To identify the origin of the flares, we analyze the $134\,\rm THz$ emissivity distribution using 3D volume rendering with magnetic field lines, alongside the corresponding NIR GRRT images for both the flaring state ($t=8,320\,GM/c^3$, total flux of $4\,\rm mJy$) and the quiescent state ($t=9,500\,GM/c^3$, total flux of $0.6\,\rm mJy$), as illustrated in Fig.~\ref{Fig: f2}. The emissivity distributions in panels (a) and (b) reveal that during flaring states, the accretion flow develops extended spiral structures of emissivity. In the zoomed-in region of panel (a) (upper left), highly twisted magnetic field lines are observed to coincide with regions of enhanced emissivity. Notably, the magnetic field lines on either side of the spiral structure exhibit opposite polarities, indicating the presence of a polarity inversion layer. This configuration strongly suggests that the high-emissivity regions are caused by energy release via magnetic reconnection events associated with these large-scale polarity inversion events. In contrast, in the quiescent state the highly twisted magnetic field lines and extended spiral-like high-emissivity structure are not clearly seen in both the volume rendering result of the GRMHD simulation and the GRRT image (see panels (b) and (c)). In this state, many small-scale polarity inversion events are seen, but they do not contribute significantly to the emission.

Panel (c) reveals a luminous extended structure (reddish features) in the GRRT image, which is not observed during the quiescent state. To identify the origin of this emission, we analyze the intensity evolution along the geodesic path of individual light rays within the GRMHD simulation domain. This is obtained by extracting ray path data (position and intensity) during GRRT post-processing. Detailed information is seen in Appendix~\ref{Sec: ray_path}. This approach enables us to quantitatively track how radiation accumulates as photons traverse polarity inversion events, thereby identifying the exact regions contributing to the observed emission. In panel (c) of the GRRT image, we highlight two positions in the most luminous region with a gold triangle (position A) and a blue star (position B), representing two distinct light rays traced during the GRRT calculations. As shown in Fig.~\ref{Fig: ray_path}, we can directly identify where each light ray acquires the majority of its emission. Once a ray exits the flaring regions, its intensity remains constant as it travels to the distant observer. 
Specifically, for position A, the emission originates from $r\approx 3.8$, $\theta \approx 138^\circ$, and $\phi \approx 212^\circ$, while for position B, it comes from $r \approx 3.0$, $\theta \approx 115^\circ$, and $ \phi \approx 321^\circ$ (with $\theta$ and $\phi$ in radians). In the 3D volume rendering image shown in Panel (a), we mark these positions with gold and blue spheres, which correspond precisely to two distinct high-emissivity regions. As these regions rotate toward the observer, whose azimuth angle is zero, Doppler boosting enhances their emission, leading to the observed flare. The blue sphere is located directly at the reconnection site and exhibits strong NIR emission in the GRRT image. This alignment suggests that the high-emissivity regions are driven by polarity inversion events.

As shown in Fig.~\ref{Fig: f2}(a), large-scale polarity inversion events produce a filamentary structure in the emissivity distribution. To precisely characterize the physical conditions within these sites, we present a 1D analysis through the current sheet in Fig.~\ref{Fig: 1D_cut}. This cut profile through the current sheet provides direct evidence for magnetic reconnection.

Fig.~\ref{Fig: 1D_cut}(a) reveals that the dominant toroidal field ($b^{\phi}$) undergoes a full polarity reversal, confirming it as the primary source of dissipated energy. In contrast, the poloidal field ($b^{\theta}$) maintains its sign and acts as a weak guide field, consistent with the global simulation results, where the ratio between the reconnecting and guiding is far less than $0.1$, suggesting a weak guiding field reconnection. Crucially, Fig.~\ref{Fig: 1D_cut}(b) quantifies the local plasma state in the upstream reconnection region, showing a high magnetization of $\sigma \approx 0.28$ and a plasma beta of $\beta \approx 2$.
These are the exact conditions where our subgrid models facilitate the acceleration of non-thermal particles. The 3D distribution of the magnetization and plasma $\beta$ is presented in Fig.~\ref{Fig: beta&sigma}(a) and (b). Our GRRT post-processing employs a variable non-thermal acceleration efficiency, $\tilde{\epsilon}$ (shown in Fig.~\ref{Fig: beta&sigma}(c)), to inject electrons into these localized, structured regions. The resulting non-thermal emission is therefore confined to these sites, contributing to the observed NIR flares. As seen in Fig.~\ref{Fig: beta&sigma}(d), while most of the domain is nearly thermal ($\kappa > 5$), the region of the polarity inversion event exhibits a low $\kappa$ index, signifying the strong non-thermal component responsible for the enhanced emission. 

\subsection{Energy Budget for NIR Synchrotron Emission}\label{sec: energy}

In this section, we demonstrate that the magnetic reconnection events identified in our GRMHD simulation provide a sufficient energy reservoir to accelerate electrons to the energies required for producing the NIR flares.
The characteristic frequency ($\nu_{\rm c}$) of synchrotron emission from an electron with Lorentz factor $\gamma_{\rm e}$ in a magnetic field of strength $B$ is given by:
\begin{equation}
\label{eq:nu_c}
\nu_c = \frac{3 e B \gamma_e^2}{4 \pi m_e c}.
\end{equation}
To produce photons in the NIR band, we take a representative wavelength for flare of $\lambda \approx 2.2 \, \mu\text{m}$, which corresponds to an observed frequency of $\nu_{\rm{obs}} \approx 1.38 \times 10^{14}$~Hz. In the reconnection regions of our simulation, the magnetic field strength is $B \gtrsim 80\,\text{G}$. Solving for the required Lorentz factor, we find:
\begin{equation}
\label{eq:gamma_e_req}
\gamma_e \approx \left( \frac{4\pi m_e c \, \nu_{\text{obs}}}{3 e B} \right)^{1/2} \approx 640.
\end{equation}
It shows that electrons must be accelerated to highly relativistic energies, with Lorentz factors on the order of $\gamma_e \sim 10^3$, to account for the observed NIR emission.

A crucial distinction must be made between the magnetization $\sigma_i$ of the bulk fluid (dominated by ions) and the magnetization $\sigma_e$ relevant to electrons. 
In our GRMHD simulation, we obtain the magnetization of the bulk fluid, where the inertia is dominated by protons. Therefore, the value of $\sigma \gtrsim 0.5$ found in our reconnection sites represents the ion magnetization ($\sigma_i$):
\begin{equation}
\label{eq:sigma_i}
\sigma_i = \frac{B^2}{4\pi \rho c^2} \approx \frac{B^2}{4\pi n_p m_p c^2} \gtrsim 0.5.
\end{equation}
However, the parameter that governs the direct magnetic acceleration of electrons is the electron magnetization ($\sigma_e$), which is defined using the electron rest-mass energy density:
\begin{equation}
\label{eq:sigma_e}
\sigma_e = \frac{B^2}{4\pi n_e m_e c^2}.
\end{equation}

Assuming a quasi-neutral electron-ion plasma where the number densities are approximately equal ($n_e \approx n_p$), the relationship between the two magnetization parameters is simply the ratio of the proton to electron mass:
\begin{equation}
\label{eq:sigma_relation}
\sigma_e = \left( \frac{m_p}{m_e} \right) \sigma_i \approx 1836 \, \sigma_i.
\end{equation}
Using our simulation result of $\sigma_i \gtrsim 0.5$, we can estimate the corresponding electron magnetization in the reconnection region:
\begin{equation}
\label{eq:sigma_e_val}
\sigma_e \approx 1836 \times 0.5 \approx 918.
\end{equation}
Therefore, the local electron magnetization is $\sigma_e \sim 10^3$, which represents the magnetic energy available per electron. This value is in the same order as the required electron kinetic energy, $\gamma_e \approx 640$.
The condition $\gamma_e \sim \sigma_e$ is satisfied. This demonstrates that the magnetic reconnection events in our GRMHD simulation are energetically capable of accelerating electrons to the Lorentz factors needed to produce the observed NIR flare. While more complex mechanisms like stochastic acceleration within the turbulent reconnection layer undoubtedly play a role in shaping the final particle distribution, this fundamental energy budget calculation confirms the physical viability of our proposed emission mechanism.

\subsection{Time delay and radio flares}\label{sec: 22GHz_flare}

One of the interesting aspects of the flaring activity of Sgr~$A^*$ is that the radio flares show delayed peaks compared to the NIR and X-ray flares \citep[e.g.,][]{Witzel2021}. Previous explanations for this time delay invoked adiabatic expansion of hot spots \citep{Xi2024, 2024ApJ...971...52M, Yusef-Zadeh2006}, but the physical mechanism driving these transient features remained unclear. 
In our previous discussion, we demonstrated the flaring characteristics of both millimeter and NIR light curves, with the latter identified as originating from polarity inversion events. Additionally, millimeter flares consistently follow the NIR flares.

At millimeter wavelengths, plasma self-absorption is significant. Polarity inversions give rise to evolving emission regions whose visibility is governed by local absorption effects.
To further investigate the mechanism behind the time delay, we analyze the flares at $43$ and $86\,\rm GHz$ ($t=8420$ and $8380\,GM/c^3$) which follow the NIR flare discussed in the section~\ref{Sec: NIR_origin}. We present their GRRT images (panels (a) and (b)) and track the intensity evolution at the peak emission pixel along the geodesics (panels (a) and (d)). In the GRRT images, the red crosses in the images mark the intensity maxima. While both frequencies show peak emission at similar projected distances from the BH ($38.2\,\rm \mu as$ at $86\,\rm GHz$ vs. $38.9\,\rm \mu as$ at $43\,\rm  GHz$), their emission originates from different regions. 
The intensity evolution in panels (c) and (d) reveals a frequency-dependent emission geometry. The bulk of the intensity for both 43\,GHz and 86\,GHz is generated in the same compact region ($r \lesssim 10\,r_{\rm g}$). However, the dominant contribution at 43\,GHz comes from a region that is modestly shifted to a larger radius and higher latitude compared to the 86\,GHz emission (panel d). This shift in the emitting region is the likely origin of the time delay. This effect is compounded by radiative transfer, as the lower frequency emission is also released from a larger radius due to the location of its frequency-dependent photosphere, which is further out in the more opaque plasma.
Although observations have reported a minor time delay of approximately 20 minutes between flares at 230\,GHz and in the NIR band \citep{Witzel2021}, a significant lag is not expected from a theoretical point of view, as the plasma is considered to be mostly optically thin at these frequencies. The presence of a delay in some events may therefore indicating additional physical complexities within the flaring mechanism or the structure of the emission region itself \citep{Jiang2024}.

Having established the outward shift of the emitting region in the millimeter band, we now examine the relationship between NIR and millimeter flares. The millimeter flares appear later than the NIR flares but exhibit a similar trend in their light curves, suggesting the same physical origin. Given the significant impact of plasma self-absorption at millimeter wavelengths, we compare multi-frequency light curves with and without self-absorption effects (Fig.~\ref{fig: f4}). When self-absorption is neglected, all radio frequencies peak simultaneously with the NIR flare. It directly links them to the same origin, which is the polarity inversion event shown in Fig.\ref{Fig: f2}. The observed time delays arise when self-absorption is included, as lower-frequency emission remains suppressed until the polarity inversion structure propagates outward to regions with lower plasma density and reduced absorption. This process shows the delayed onset of millimeter flares, while strengthening their connection to the NIR flares. Ultimately, our results indicate that NIR and millimeter flares originate from the same underlying polarity inversion event, with their temporal offsets governed by the frequency-dependent plasma self-absorption and emission region evolution.

\section{Summary and Discussion}

We have performed 3D two-temperature GRMHD simulations of magnetized accretion flows onto a rotating BH with multiple magnetic loops and GRRT calculations of synchrotron emission with hybrid thermal and non-thermal eDFs. Our analysis focuses on the origin of the flares of Sgr~A$^*$ at multiple frequencies and the time delays between them. Below, we summarize our conclusions point-wise:

\begin{enumerate}
    \item Our study shows the importance of non-thermal electrons to explain the observed NIR light curves of Sgr~A$^*$. The large-scale polarity inversion events generated from a multi-loop magnetic configuration in SANE accretion flows can produce NIR flares with comparable intensity and variability to observations.
    
    \item We see filament structures in GRRT images during NIR flares. By tracing the trajectories and intensity evolutions of the light rays, we confirm that the most luminous regions in GRRT images originate from the large-scale polarity inversion sites near the BH.
    
    \item The multi-frequency GRRT light curves exhibit time delays. The millimeter radio flares appear later than the NIR flares. The millimeter radio flares can lag behind NIR flares up to approximately 50 minutes, falling within the range reported by observations.

    \item The radio flares are found to have the same origin as the NIR flares, which are related to large-scale polarity inversion events. The time delay of radio flares arises from the frequency-dependent self-absorption of plasma at millimeter wavelengths. The emission region shifts outward before becoming visible, introducing a delay in the observed light curves.
   
\end{enumerate}

This paper presents large-scale polarity inversions in SANE accretion flows as a model for the NIR flares of Sgr~A$^*$, showing good agreement with current observations. This offers an alternative model to the widely studied MAD accretion flows around Sgr~A$^*$ \citep[e.g., ][]{Dexter2020, Porth2021, 2022MNRAS.511.3536S, 2023arXiv230816740N, 2024arXiv240410982A}. Together with the flares, we have shown that our model is better at explaining other observational features of the galactic center. For example, it does not have a persistent relativistic jet (see more discussion about the accretion states and jet in Appendix~\ref{sec: jet_nojet}), it predicts time delays between near-infrared and 43 GHz flares, and it explains the observed SED.

However, the multi-loop configuration still has limitations. It cannot maintain frequent, long-term, large-scale polarity inversion events. As previous works indicate, it is more akin to a transition phase \citep{2023MNRAS.522.2307J, Jiang_2024}. How persistent flaring events can be created for Sgr~A$^*$ is still a question to address in the future.
While it is sufficient to capture the formation and large-scale dynamics of magnetic polarity inversions, the resolution employed in this work is not intended to fully resolve the small-scale turbulence or the asymptotic, plasmoid-dominated regime of magnetic reconnection. We should note that the studies using ultra-high resolutions by GPU-accelerated GRMHD codes \citep{Ripperda2021} reach the asymptotic regime where reconnection properties become independent of the grid requires computational resources. It is beyond the scope of this global 3D study due to limited computational resources. Therefore, our analysis focuses on the large-scale polarity inversion processes rather than the microphysical details of the reconnection layer itself.

Due to limited numerical resolution, the mean-field dynamo in the accretion disk is not fully resolved, which did not persistently generate alternating polarities of the magnetic field \citep{DelZanna2022, Mattia2020, Mattia2022}. We also cannot induce the tearing instability required for the formation of plasmoids at the reconnection site \citep[e.g., ][]{Ripperda2021,2023MNRAS.522.2307J}, which may reduce the amount of non-thermal electrons.
For a polarity inversion event to be observed as a flare, two key conditions must be met. First, the event must be intrinsically energetic, as most are too weak to produce a noticeable effect. Second, the resulting plasma must be moving towards the observer at relativistic speeds, causing its emission to be significantly amplified by Doppler boosting. Conversely, an equally powerful event with plasma moving away from our line of sight would be de-boosted and appear faint. These observational biases, therefore, impact both the number and the apparent magnitude of the flares we detect.

Our GRRT code cannot currently incorporate Compton scattering of photons, meaning our model cannot explain X-ray counterparts.
The synchrotron self-Compton process may also be responsible for the abundance of X-rays in the flaring events \citep[e.g.,][]{2012A&A...537A..52E, 2016MNRAS.461..552D}, which we neglect in this work. Our ray-tracing technique also has limitations on extracting the local property of the flaring region, which can be approached from a Monte Carlo method by tracing photons from a reconnection event to infinity. We will investigate these missing processes by using higher-resolution simulations to resolve the lower flux of X-ray flares in future work. 

\begin{acknowledgments}
We greatly thank Dr. Ben S. Prather for his great support of the KHARMA code and thank Xufan Hu for useful discussions regarding this work. We appreciate the constructive comments provided by the anonymous reviewers, which have improved the manuscript.
This research is supported by the National Key R\&D Program of China (No.~2023YFE0101200), the National Natural Science Foundation of China (Grant No.~12273022, 12192220, 12133008), and the Shanghai Municipality orientation program of Basic Research for International Scientists (Grant No.~22JC1410600). The simulations were performed on TDLI-Astro, Pi2.0, and Siyuan Mark-I at Shanghai Jiao Tong University.

\end{acknowledgments}

%




\newpage
\appendix
\section{Code setup}
\subsection{GRMHD simulation setup {\rm KHARMA}} \label{Sec: kharma}
Our GRMHD simulations are implemented using the KHARMA\footnote{\url https://github.com/AFD-Illinois/kharma} code \citep{2024arXiv240801361P}, which is a GPU-accelerated version of {\tt iharm3D} \citep{2021JOSS....6.3336P}.
Both KHARMA and {\tt iharm3D} codes are based on the {\tt HARM} code \citep{2003ApJ...589..444G}, which solves the ideal MHD equations in the framework of general relativity (GR). The ideal GRMHD equations are solved for an Eulerian observer and are written as follows:
\begin{equation}
    \begin{aligned}
        \partial_t(\sqrt{-g}\rho u^t)&=-\partial_i(\sqrt{-g}\rho u^i), \\
        \partial_t(\sqrt{-g}T^t_{\nu})&=-\partial_i(\sqrt{-g} T^i_{\nu}) + \sqrt{-g}T^\kappa_\lambda\Gamma^\lambda_{\nu\kappa}, \\
        \partial_t(\sqrt{-g}B^i)&=-\partial_j\left[\sqrt{-g}(b^ju^i - b^i u^j)\right], \\
        \frac{1}{\sqrt{-g}}\partial_i(\sqrt{-g}B^i)&=0,
    \end{aligned}
\end{equation}
where $\rho$ is the rest mass density, $u^\mu$ is the four-velocity, $\Gamma^\alpha_{\beta \gamma}$ is the connection coefficient, $B^i$ and $b^\mu$ are the three and four-magnetic field components, $T^\mu_\nu$ energy momentum tensor, and $g$ is the metric determinant \citep[for details see][]{2003ApJ...589..444G}. KHARMA provides a versatile, mobile, and adaptable approach that fits well across various applications. This is achieved by adopting the Parthenon Adaptive Mesh Refinement Framework \citep{2022arXiv220212309G} and the Kokkos programming module \citep{9485033}.  Additionally, we implement a floor model to ensure that the GRMHD code can handle the highly magnetized low-density regions, particularly close to the BH and the rotation axis \citep{Rezzolla-Zanotti2013}. Specifically, we set the floor rest mass density and pressure as $\rho_{\rm fl}=10^{-6}\, r^{-3/2}$ and $p_{\rm fl}=10^{-8}\,r^{-5/2}$, respectively (reference). Also, we set the maximum values of magnetization and Lorentz factor as $\sigma_{\rm max}=50$ and $\gamma_{\rm max}=15$, respectively. Finally, to ensure divergence-free magnetic field throughout the simulation domain, we use the Face-Centered Constrained Transport (Face-CT) method \citep{2024arXiv240801361P}. Specifically, we use the gs05$_c$ Face-CT scheme, which corresponds to the upwind method described in \cite{2005JCoPh.205..509G} ($E_z^c$ method, Eqs. 42 \& 51), extended to 3D in \cite{2008JCoPh.227.4123G}.

\begin{figure}
\centering
 	\includegraphics[height=.6\linewidth]{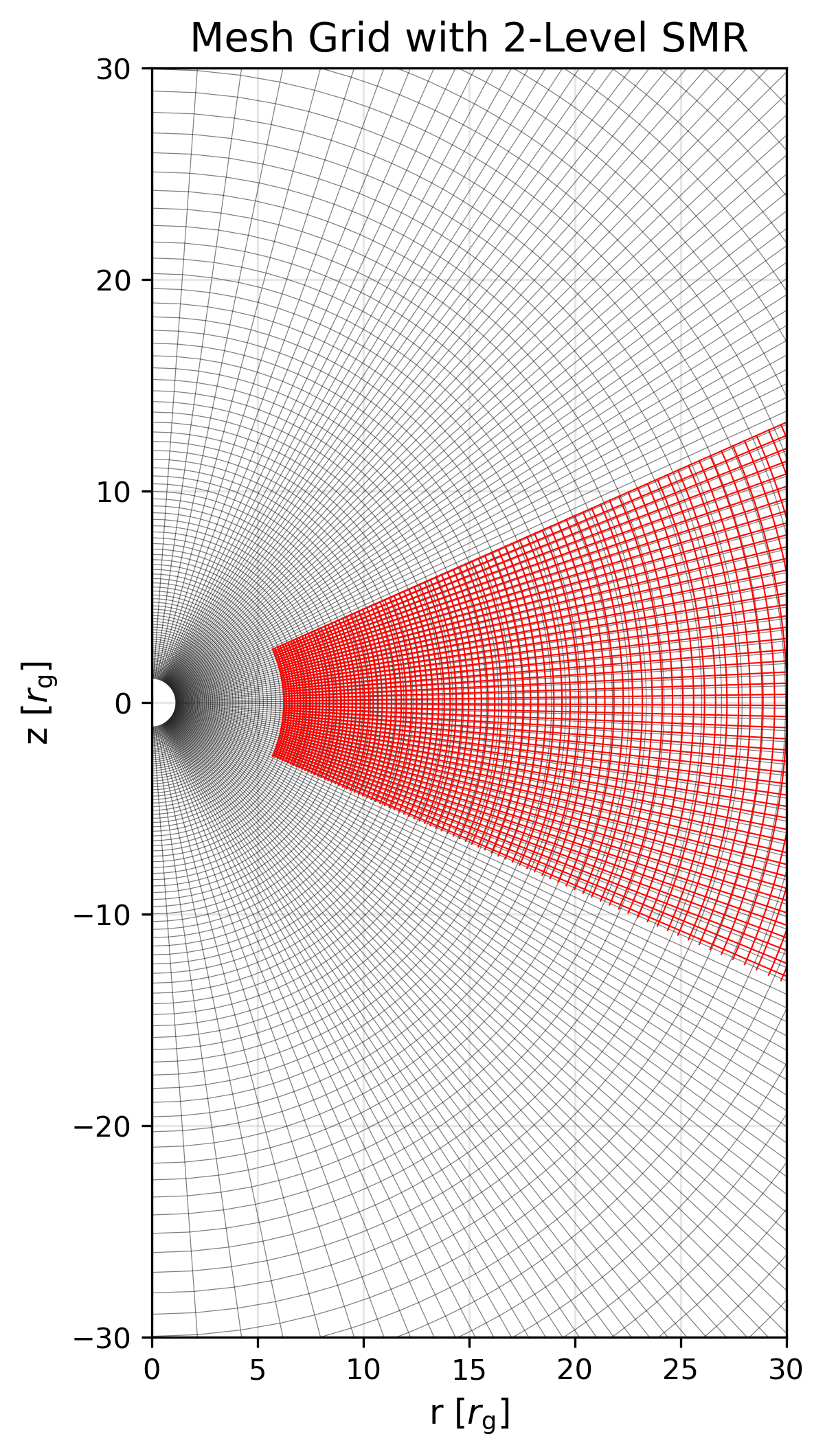}
        \includegraphics[height=.6\linewidth]{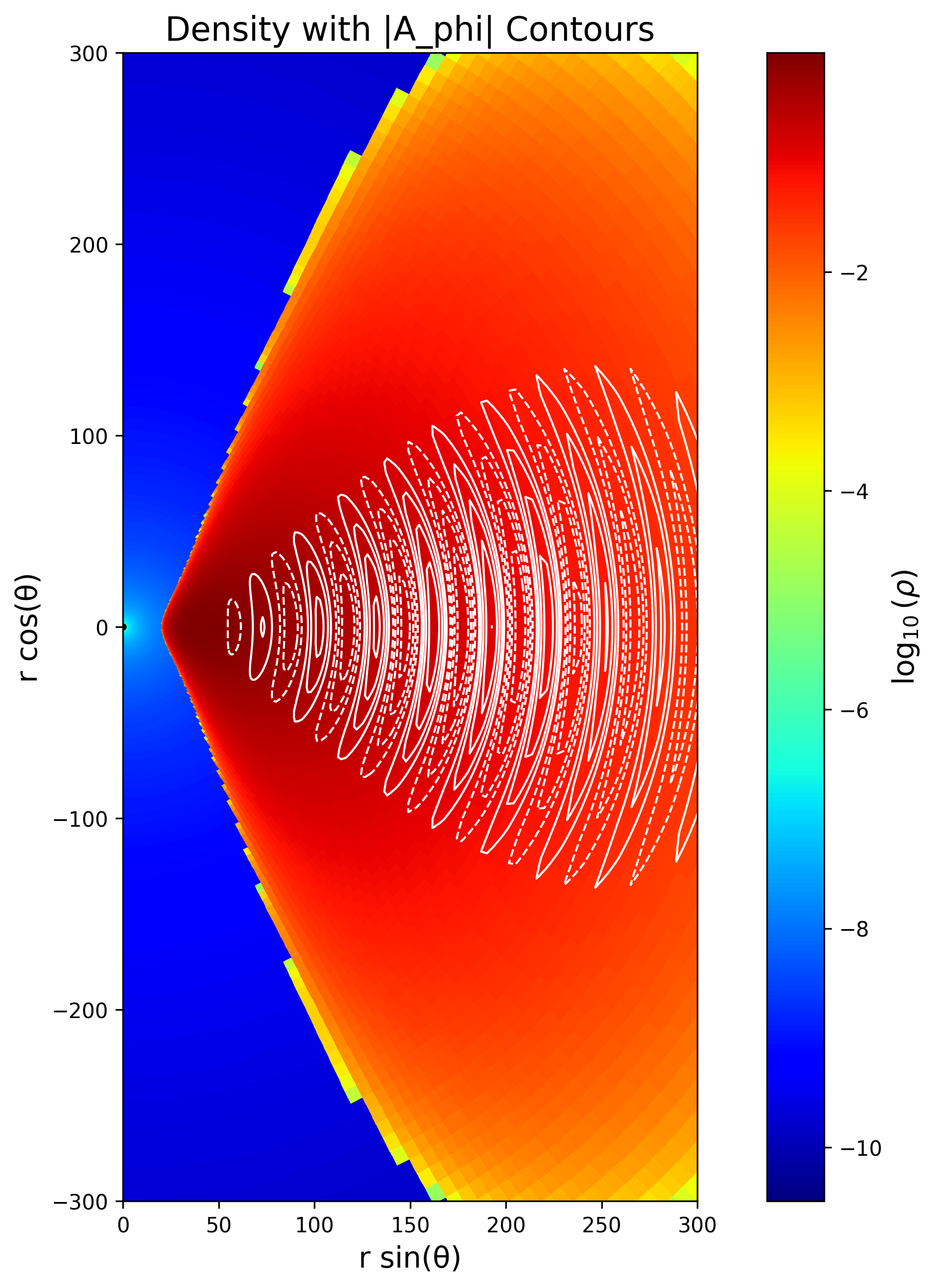}
 	\caption{Computational grid distribution ({\it left}) and initial magnetic configuration ({\it right}). In the {\it left} panel, we use 2-level SMR. Based grid and refined grid regions are presented in black and red, respectively.
    In the {\it right} panel, we show the initial magnetic configuration using the contours of the vector potential $A_\phi$, the solid and dashed contours represent the polarity of the loops.}
 	\label{Fig: mesh}
\end{figure}

\subsection{Resolution and initial magnetic configuration} \label{sec:res_check}

The simulation utilizes a base resolution of \(192 \times 144 \times 128\) with 2 levels of static mesh refinement (SMR), achieving an effective resolution of \(384 \times 288 \times 256\). The second level of SMR is applied within a specific region defined in the Wide-pole Kerr-Schild (WKS) coordinates (details see \citep{2024ApJ...977..200C}, a similar approach also seen \cite{2006MNRAS.368.1561M,2012MNRAS.423.3083M,2017MNRAS.467.3604R}). The radial extent of this region spans from $r_{\text{min}} = 6.24\,r_{\rm g}$ to $r_{\text{max}} = 281\,r_{\rm g}$, while the polar angle ranges from $\theta_{\text{min}} = 66^\circ$ to $\theta_{\text{max}} = 114^\circ$. The grid configuration of the simulation is plotted in the left panel of Fig.~\ref{Fig: mesh}, with the red part representing the refined region.
 
In the right panel of Fig.~\ref{Fig: mesh}, we present the initial multi-loop magnetic configuration with magnetic loops along the radial direction with alternating polarities.

\subsection{Cooling effect} \label{sec: cooling}

Cooling has a strong impact on the particle distribution function, which limits the maximum Lorentz factor and changes the power-law index in the distribution function. In \cite{2022MNRAS.511.3536S}, a simple approximation was proposed to mimic the effect of cooling. The idea is written as follows:

The synchrotron cooling time $t_{\rm sync}$ is
\begin{equation}
    t_{\rm sync} = \frac{3m_{\rm e}c}{4\sigma_{T}U_{\rm B}\gamma\beta^2},
\end{equation}
where $\sigma_{\rm T}$ is the Thomson cross-section, $U_{\rm B}$ is the magnetic energy density, $\gamma$ is the Lorentz factor and $\beta\equiv v/c\sim1$. The escape time scale for the synchrotron photon is roughly the light crossing time $t_{\rm esc}=r_{\rm g}/c$. Therefore, cooling becomes important around
\begin{equation}
    \gamma_{\rm break} \approx 3.9\times 10^{3}\left(\frac{|b|}{100\,\rm Gauss}\right)^{-2},
\end{equation}
which results in a typical break frequency of
\begin{equation}
    \nu_{\rm break}\approx 2.5\times 10^{15}\left(\frac{|b|}{100\,\rm Gauss} \right)^{-1}\,\rm Hz.
\end{equation}
It leads to a strong cooling effect at frequencies higher than $\sim 10^{15}\,\rm Hz$. In practice, following \cite{2022MNRAS.511.3536S}, when $\nu>\nu_{\rm break}$, we divide the calculation of emissivity and absorptivity in Eq.~\ref{Eq: emi&alpha} by a factor $\sqrt{\nu/\nu_{\rm break}}$. With such a cooling effect, the X-ray emission is greatly reduced, providing a more reasonable SED at the X-ray band.

\section{Supplementary information}
\subsection{Spectral Energy Distribution} \label{sec: sed}

\begin{figure*}
    \centering
    \includegraphics[width=.7\linewidth]{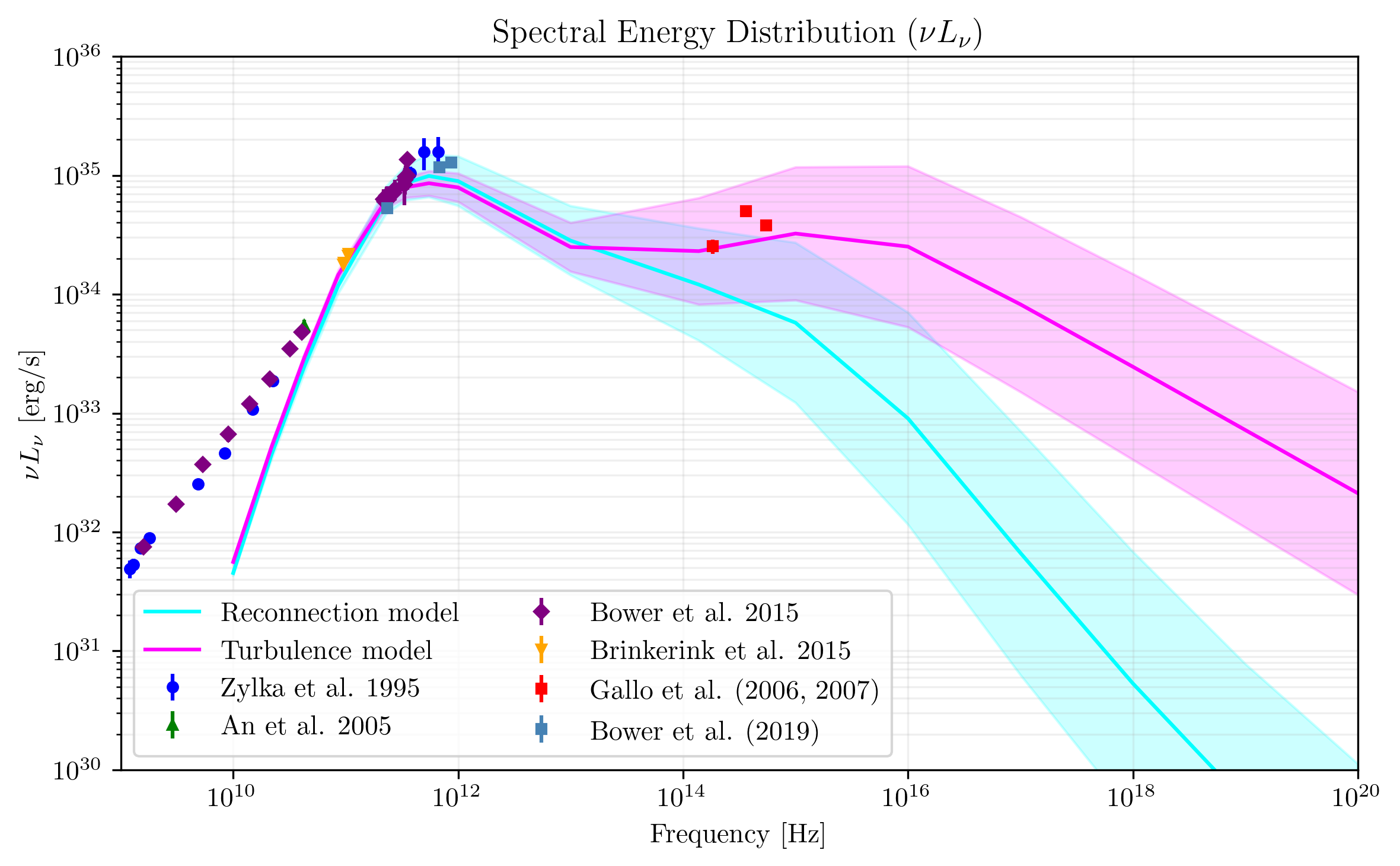}
    \caption{Spectral energy distribution from observations and simulations. The observational data is compiled from \cite{1995A&A...297...83Z, 2005ApJ...634L..49A, 2015ApJ...802...69B, 2015ASPC..499..167B, 2019ApJ...881L...2B, 2017MNRAS.466.4121C, 2006MNRAS.370.1351G, 2007ApJ...670..600G}. The pink and cyan solid lines represent the averaged SED from our turbulence and reconnection models, respectively, with the averaging performed over $t = 8,000$ to $11,000\,\rm [GM/c^3]$. The shaded region indicates the standard deviation of the SED within this time range.}
    \label{fig: SED}
\end{figure*}

In this subsection, we present a comparative analysis of the spectral energy distribution (SED) between observational data (millimeter to submillimeter data comes from \cite{1995A&A...297...83Z, 2005ApJ...634L..49A, 2015ApJ...802...69B, 2015ASPC..499..167B, 2019ApJ...881L...2B}, and IR data comes from the collection in \cite{2017MNRAS.466.4121C} originally from \cite{2006MNRAS.370.1351G, 2007ApJ...670..600G}) and modelling results from our simulations, shown in Fig.~\ref{fig: SED}.

In this work, we employed two distinct subgrid models to calculate electron temperature and non-thermal emission (see details in Sec.~\ref{sec:2T}). These models incorporate the microphysics of turbulent and reconnecting plasma, enabling us to apply this information in our GRMHD and GRRT simulations without resolving the detailed structures of turbulence and reconnection. However, identifying these turbulence and reconnection structures and applying the appropriate subgrid models to those regions is a challenge. In this study, we do not attempt to address this challenge but instead adopt similar methods in previous works \citep[e.g., ][]{Fromm2022, Yang2024} to investigate flares. 

In addition to the results based on the turbulence subgrid model discussed in the main text, we add the reconnection model for comparison in Fig.~\ref{fig: SED}. We plot the averaged SED within the range $t=8,000-11,000\,GM/c^3$ for both of models in cyan and pink, with shaded region represents the standard deviation. As our previous studies \citep{2023MNRAS.522.2307J, Jiang_2024} have demonstrated both electron heating models produce similar results at millimeter to submillimeter frequencies, where thermal emission predominates. At higher frequencies, non-thermal emission becomes dominant. The two subgrid models for non-thermal emission differ in the amounts of energy injected into the non-thermal electrons, leading to distinct outcomes at the high-frequency end of the SED. In the IR band, the turbulence model has better agreement with observation and generates higher NIR flux during flares that are closer to the observed peak flux in \cite{Abuter2018} ($\sim10-25\,\rm mJy$). Therefore, we make it a preferred choice over the reconnection model the discussion of this work.

Major discrepancy between our simulations and observations appears at frequencies lower than 43 GHz, where our simulations underpredict the radio emission. Since in GRMHD simulations, the jets cool adiabatically as they expand, reducing electron temperature and leading to insufficient low-frequency emission compared to the observed flat spectrum, which suggests an isothermal jet. In our multi-loop model, polarity inversion prevents the formation of a strong jet, further contributes to this underprediction due to the lack of jet activity. 

\subsection{Ray path and intensity from GRRT calculations}\label{Sec: ray_path}

\begin{figure}
\centering
 	\includegraphics[height=.3\linewidth]{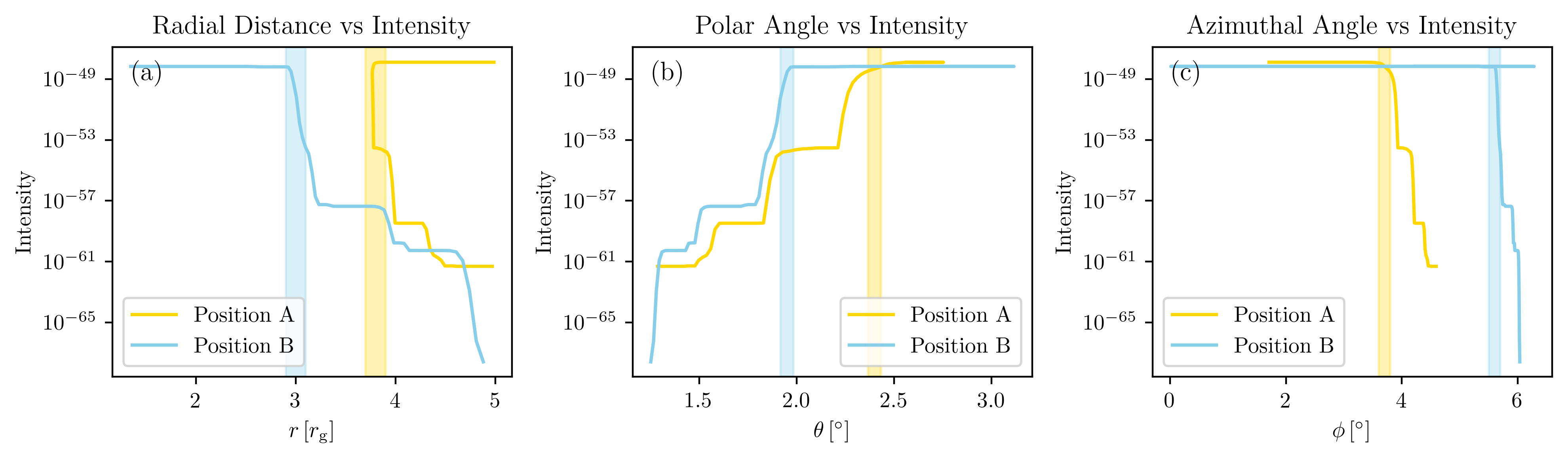}
 	\includegraphics[height=.4\linewidth]{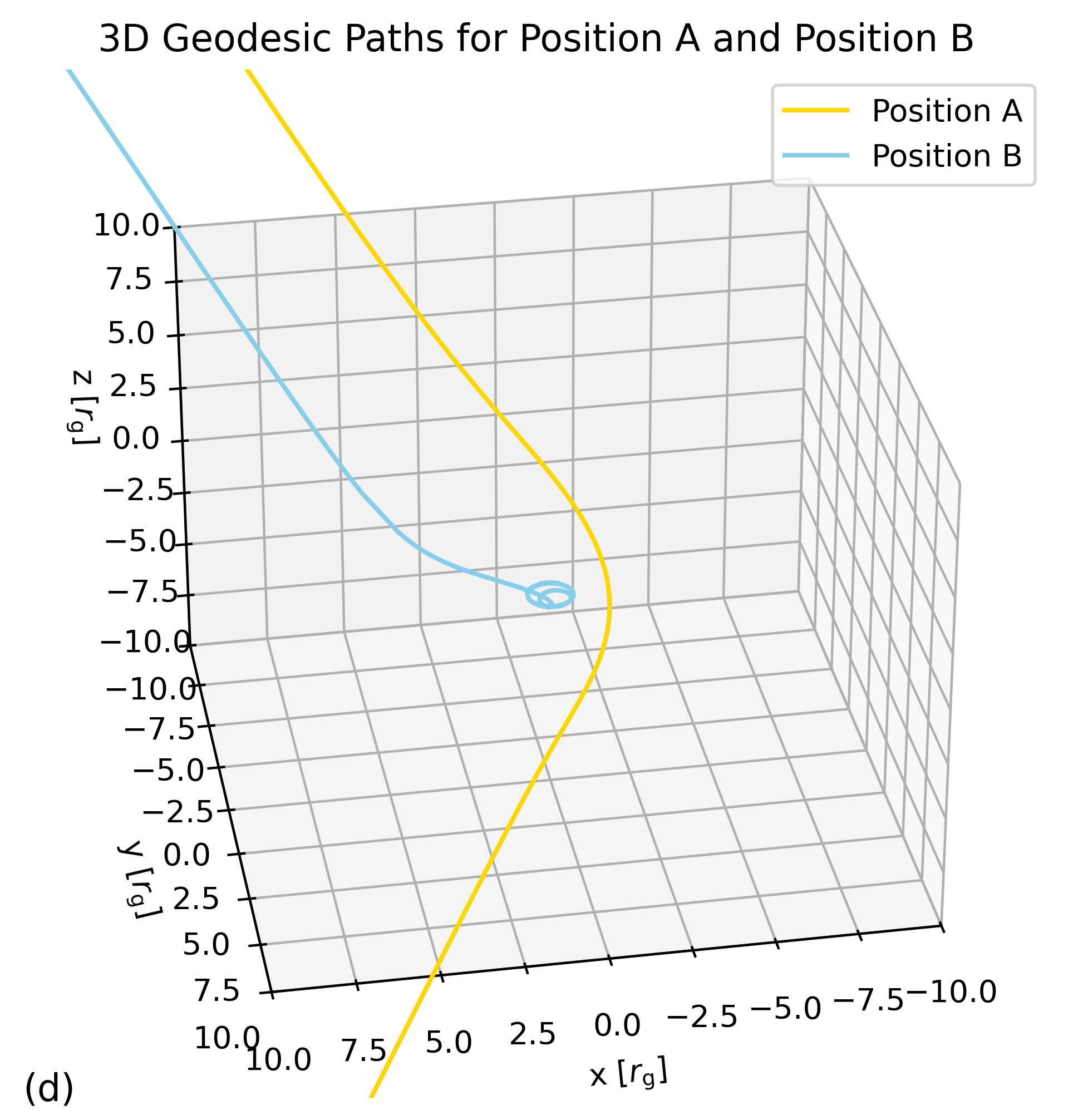}
 	\caption{Geodesic line and intensity evolution. Panels (a) to (c) show the intensity evolution along the ray paths, with each panel representing a different coordinate direction: $r, \theta$, and $\phi$. The yellow and blue lines correspond to the two positions marked in the GRRT image in Panel (c) of Fig.~\ref{Fig: f2}. In Panel (d), the 3D geodesic trajectories of these two rays are presented, serving as a spatial reference for understanding the paths along which the intensity evolves in panels (a)-(c).}
 	\label{Fig: ray_path}
\end{figure}

To track the intensity accumulation process during the GRRT calculation for each pixel, we extract geodesic line data along with the corresponding light ray intensity at each position. In Fig.~\ref{Fig: ray_path}, we show the intensity evolution in spherical coordinates, allowing us to identify where the ray acquires most of its emission and determine the exact location of the flares.

In this figure, the two rays corresponding to Positions A and B in the GRRT image (Panel (c) of Fig.\ref{Fig: f2}) are plotted in yellow and blue, respectively, with the shaded region indicating the flare coordinates where most of the emission is gained. This enables us to directly locate the corresponding positions in the 3D volume-rendered image shown in Fig.\ref{Fig: f2}.

In Panel (d) of Fig.~\ref{Fig: ray_path}, we present the paths of these two rays, illustrating our ray-tracing trajectories.

\subsection{Cross-correlation function} \label{Sec: CCF}

The cross-correlation function (CCF) is a statistical tool used to measure the time lag between two temporal signals \citep{1997ASSL..218..163A}. In practice, the discrete cross-correlation function (DCCF) is more widely used instead of the continuous CCF. Given that two light curves $F(t)$ and $G(t)$, for each fair of data point $F(t_j)$ and $G(t_k)$, the time lag between them is
\begin{equation}
  \Delta t_{jk} = t_j-t_k.  
\end{equation}
Constructing a list of time lag $\tau$, within each time lag bin $\tau_i\leq \Delta t_{jk} \leq \tau_{i+1}$, the correlation coefficient for each pair is given by
\begin{equation}
    U_{jk} = \frac{(F(t_j)-\overline{F})(G(t_k) - \overline{G})}{\sigma_F\sigma_G},
\end{equation}
where $\overline{F}$ and $\overline{G}$ are the averaged values of the two light curves, and $\sigma_F$ and $\sigma_G$ are their standard deviations. Therefore, the DCCF within the given bin of time lag $\tau_i$ is obtained as 
\begin{equation}
    DCCF(\tau_i) = \frac{1}{n_i}\sum_{jk} U_{jk},
\end{equation}
where $n_i$ is the number of pairs within the bin $\tau_i$. The error of DCCF can be measured as the standard deviation within each bin:
\begin{equation}
    \sigma_{DCCF}(\tau_i) = \frac{1}{\sqrt{n_i}}\sqrt{\sum_{jk}\left(U_{jk}-DCCF(\tau_i)\right)^2}.
\end{equation}

\subsection{Different stages of accretion flow from multi-loop magnetic configuration and convergence check} \label{sec: jet_nojet}

\begin{figure}
    \centering
    \includegraphics[width=\linewidth]{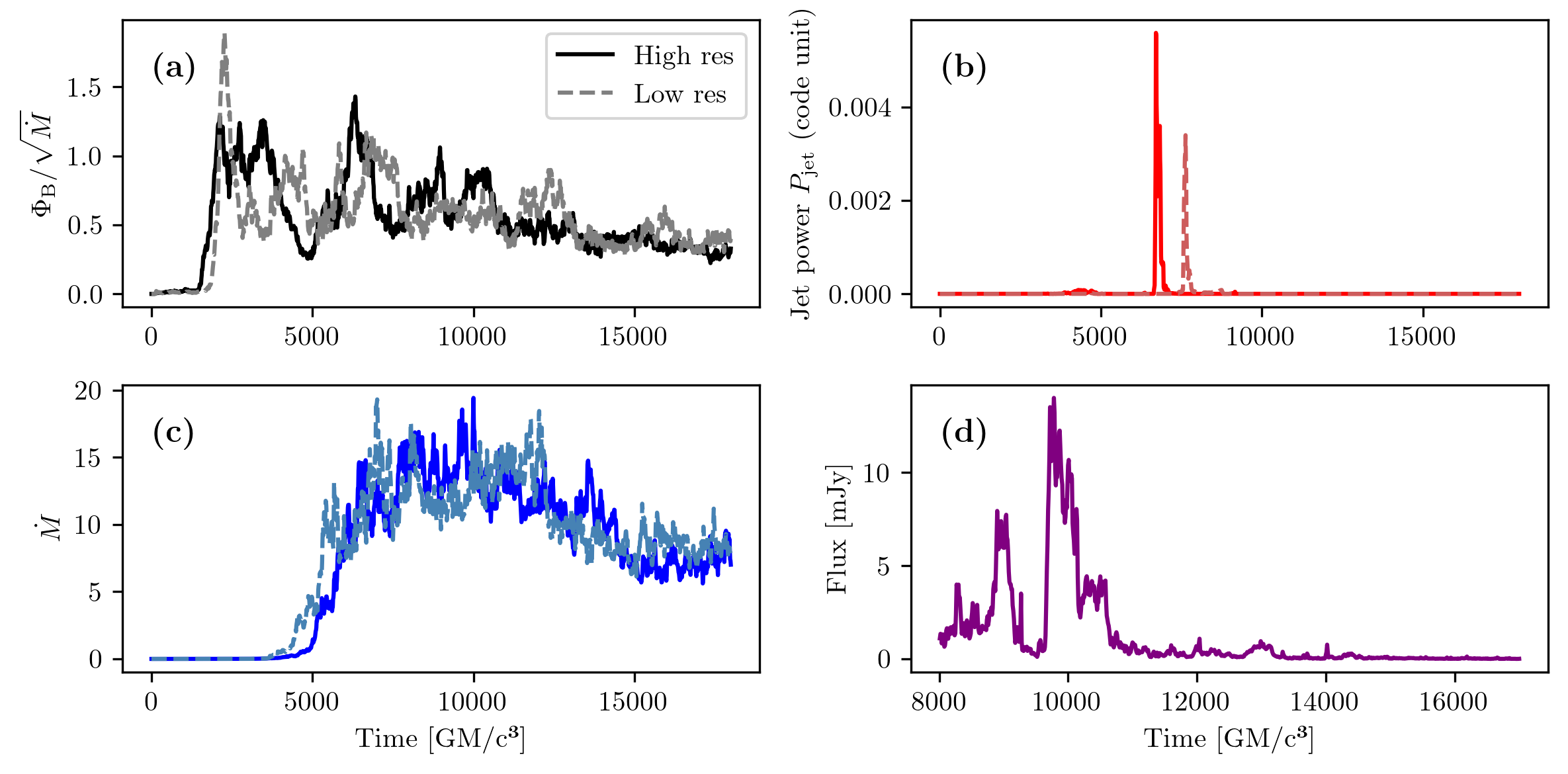}
    \caption{Panels (a)-(c) show the time evolution of the normalized magnetic flux (black), jet power (red), and accretion rate (blue). Solid lines represent the fiducial run, while dashed lines indicate the low-resolution test. Panel (d) presents the NIR light curve from $t = 8,000$ to $18,000\,GM/c^3$ (purple).}
    \label{fig: Mdot}
\end{figure}
In Fig.~\ref{fig: Mdot}, we present the time evolution of the magnetic flux, jet power, and accretion rate, following the definitions of $\dot{M}$ and $\Phi_{\rm B}$ from \cite{Porth2019} and $P_{\rm jet}$ from \cite{Nathanail2020} in Panel (a)-(c). The magnetic flux $\Phi_{\rm B}$ and mass accretion rate $\dot{M}$ is calculated at the event horizon given by \citep{Porth2019}: $\Phi_{\rm B} = \frac{1}{2} \int_0^{2\pi} \int_{0}^{\pi} \left|B^r\right| \sqrt{-g} \, d\theta \, d\phi$, $\dot{M} = \int_0^{2\pi} \int_0^{\pi} \rho u^r \sqrt{-g} \, d\theta \, d\phi$. The jet power, following \cite{Nathanail2020}, is given by $P_{\rm jet} = \int_0^{2\pi} \int_{\theta_{\rm jet}} (-T_t^r - \rho u^r) \sqrt{-g} \, d\theta \, d\phi$ at $r=50\,r_{\rm g}$
where the integration range $\theta_{\rm jet}$ covers the funnel region ($\sigma \geq 1$). 
From the evolution of the accretion rate and magnetic flux, we can identify a phase transition in the accretion flow.
The simulation is performed for more than $t=18,000\,GM/c^3$. After $~8,000\,G M/c^3$ the inner part of the accretion flow ($\lesssim 50\,r_{\rm g}$) reaches a quasi-steady state. Before $t = 12,000 \,GM/c^3$, the flow undergoes frequent polarity inversion events, resulting in more variable $\dot{M}$ and $\Phi_{\rm B}$. The polarity inversion events also drive the flaring events seen in the NIR light curve in Panel (d). After this period, as magnetic energy dissipates and polarity inversions cease, the NIR light curve transitions into a quiescent state. The frequent magnetic reconnection during polarity inversion events suppresses jet launching, leading to a near-zero jet power for most of the time, indicating the absence of a persistent jet seen in Panel (b).

From Panels (a) to (c), we present results from two GRMHD simulations, one with SMR and one without, corresponding to resolutions of $384\times288\times256$ and $192\times144\times128$, respectively. The similar evolution observed in these panels between the two resolutions suggests that our simulations have achieved numerical convergence.

\subsection{Impact of $\sigma_{\rm cut}$}
\label{sec: sigma_cut}
\begin{figure}
\centering
 	\includegraphics[width=.8\linewidth]{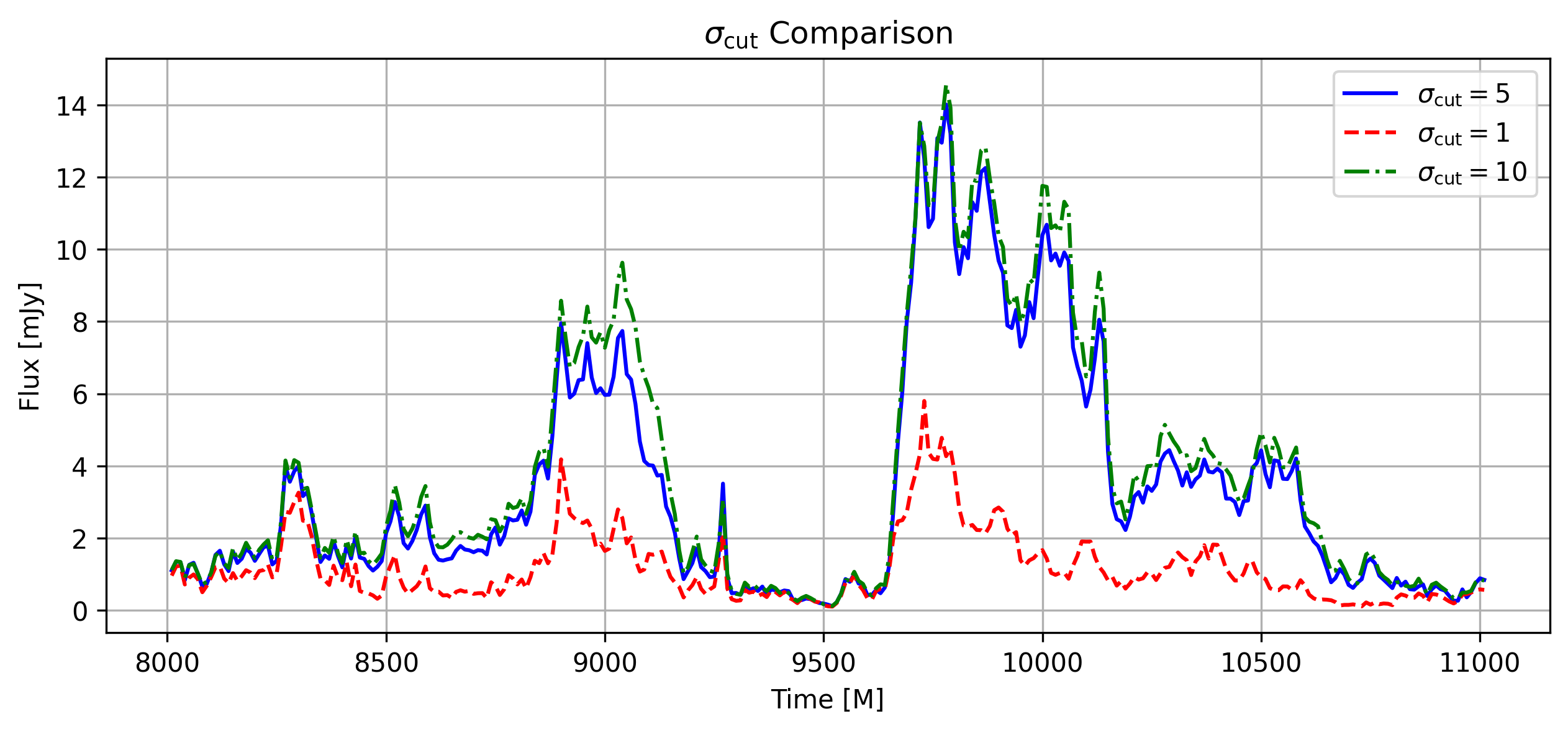}
 	\caption{Comparison of NIR light curves for different $\sigma_{\rm cut}$ values. The solid blue line represents the fiducial run ($\sigma_{\rm cut} = 5$), the dashed red line corresponds to $\sigma_{\rm cut} = 1$, and the dot-dashed green line represents $\sigma_{\rm cut} = 10$.}
 	\label{Fig: sigma_cut_comp}
\end{figure}

In GRRT post-processing, highly magnetized regions in the GRMHD data are excluded based on the parameter $\sigma_{\rm cut}$, which may impact the NIR light curves. In this section, we analyze the effect of different values of $\sigma_{\rm cut}$ by comparing the NIR light curves for $\sigma_{\rm cut}= 1$ and $\sigma_{\rm cut} = 10$ against our fiducial choice of $\sigma_{\rm cut} = 5$, as shown in Fig.~\ref{Fig: sigma_cut_comp}.

Our results indicate that for $\sigma_{\rm cut} = 1$, the light curve exhibits a noticeable decrease. However, when $\sigma_{\rm cut}$ increases to 10, the difference between the cases with $\sigma_{\rm cut} = 10$ and $\sigma_{\rm cut} = 5$ becomes negligible. This suggests that our fiducial choice of $\sigma_{\rm cut} = 5$ is already sufficient, and further increasing this parameter does not affect the GRRT calculation results.

\begin{figure}
\centering
 	\includegraphics[width=\linewidth]{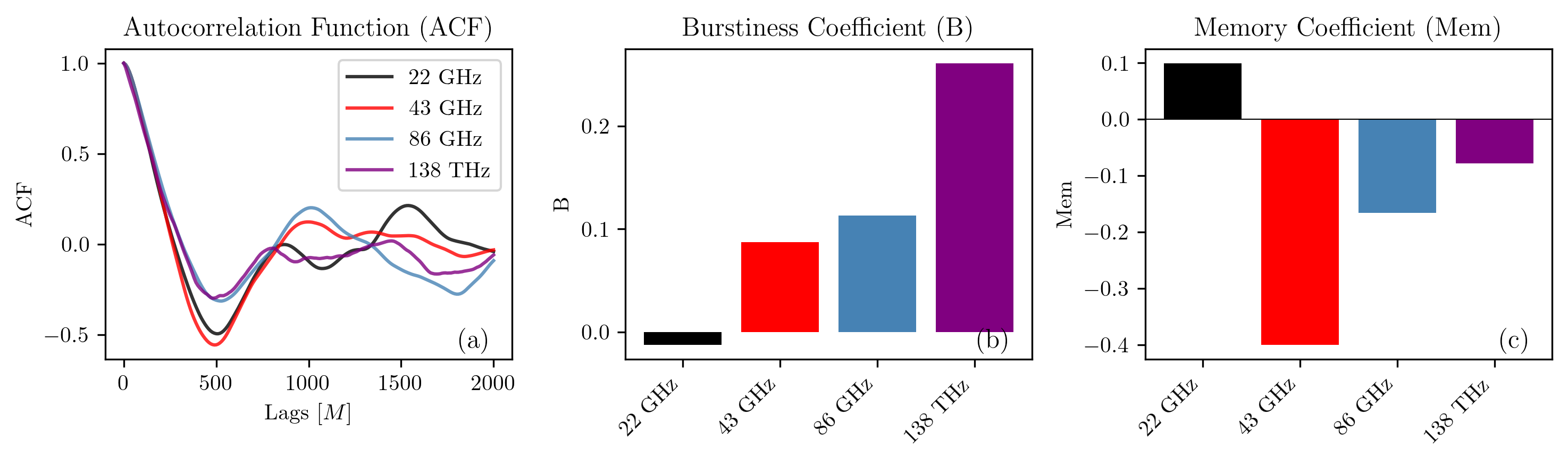}
 	\includegraphics[width=.45\linewidth]{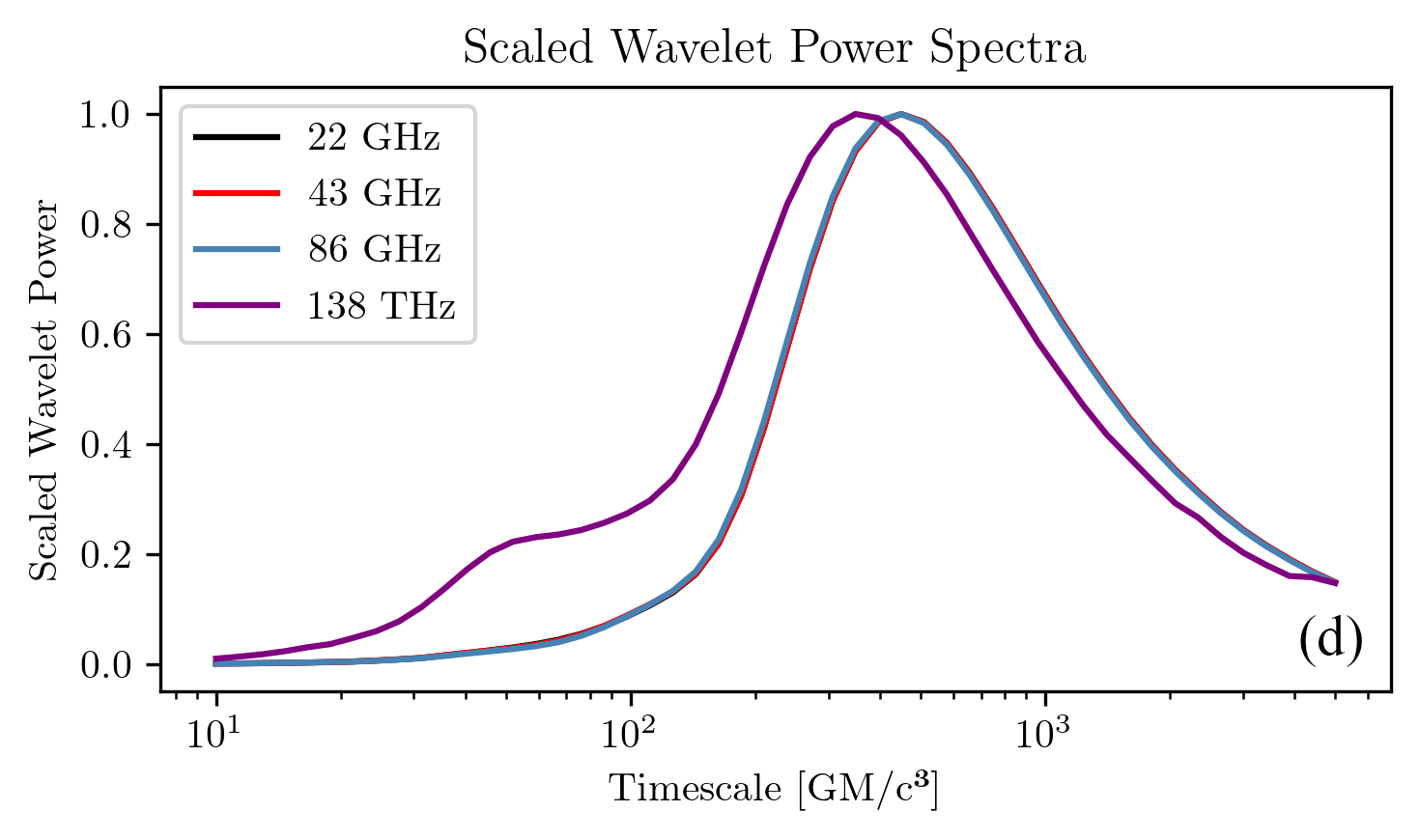}
 	\caption{Statistical analysis of the multi-frequency light curves. (a) Autocorrelation functions of the light curves at different frequencies: $22,\rm GHz$ (black), $43,\rm GHz$ (red), $86,\rm GHz$ (blue), and $138,\rm THz$ (purple). The same color scheme is used in Panels (b)-(d). (b) and (c) display the burstiness and memory coefficients, while (d) presents the scaled wavelet power spectra.}
 	\label{Fig: ACF}
\end{figure}

\subsection{Further statistics analysis on the light curves}
\label{sec: statistics}

In this section, we present more detailed statistics on the light curves to identify the periodicities and properties across them. 

The multi-frequency light curves in Fig.~\ref{fig: f1} show systematic frequency-dependent variations in temporal properties, suggesting different physical processes in the source. The autocorrelation function (ACF) (definition see \cite{1976tsaf.conf.....B, 2019A&A...627A.103V}) in Panel (a) of Fig.~\ref{Fig: ACF} shows rapid decorrelation at $138\,\rm THz$ (short-timescale variability), while radio frequencies ($22-86 \,\rm GHz$) retain correlations over longer lags, suggesting a more periodic behavior. In Panel (b), the burstiness coefficient (definition see \cite{2019PhRvE.100b2307J}) of the 4 light curves is presented, showing an increasing trend as frequency increases. It indicates more clustered, stochastic variability at higher frequencies compared to smoother variations at lower frequencies. The memory coefficient (definition see \cite{2019PhRvE.100b2307J}) shown in Panel (c) transitions from weakly positive (persistent trends) at $22\,\rm GHz$ to negative (anti-correlated behavior) at intermediate frequencies (43-86 GHz), with a slight rebound toward less negative values at $138\,\rm THz$. These trends imply shorter flare durations and intervals at higher frequencies, where compact, turbulent regions dominate. Lower frequencies, in contrast, trace larger, more stable structures with slower variability.

Similarly, the scaled wavelet power spectra in Panel (d) of Fig.~\ref{Fig: ACF} show overlapping radio curves, indicating similar variability patterns across these frequencies, with a peak at $\sim 450\,GM/c^3$. In contrast, the $138\,\rm THz$ spectrum exhibits a major peak at $\sim 348\,GM/c^3$ and a smaller bump at $\sim 80\,GM/c^3$. This suggests that NIR flares have shorter durations or smaller intervals, likely coming from the rapid orbital motion in the innermost regions ($3$-$5\,r_{\rm g}$) at $\sim 0.3c$ (see similar structure in Panel (a) of Fig.~\ref{Fig: f2}). The high-frequency structure in the spectra corresponds to the small spikes seen in NIR light curves during flares.


\bibliography{sample631}{}
\bibliographystyle{aasjournal}



\end{document}